\documentclass{article}

\usepackage{arxiv}

\usepackage[utf8]{inputenc} 
\usepackage[T1]{fontenc}    
\usepackage{lmodern}        
\usepackage{hyperref}       
\usepackage{url}            
\usepackage{booktabs}       
\usepackage{amsfonts}       
\usepackage{nicefrac}       
\usepackage{microtype}      
\usepackage{graphicx}

\title{The harms of class imbalance corrections for machine learning
based prediction models: a simulation study.}

\author{
    Alex Carriero
   \\
    Julius Center for Health Sciences and Primary Care \\
    University Medical Center Utrecht \\
  Netherlands \\
  \texttt{\href{mailto:a.j.carriero@umcutrecht.nl}{\nolinkurl{a.j.carriero@umcutrecht.nl}}} \\
   \And
    Kim Luijken
   \\
    Julius Center for Health Sciences and Primary Care \\
    University Medical Center Utrecht \\
  Netherlands \\
  \texttt{} \\
   \And
    Anne de Hond
   \\
    Julius Center for Health Sciences and Primary Care \\
    University Medical Center Utrecht \\
  Netherlands \\
  \texttt{} \\
   \And
    Karel GM Moons
   \\
    Julius Center for Health Sciences and Primary Care \\
    University Medical Center Utrecht \\
  Netherlands \\
  \texttt{} \\
   \And
    Ben van Calster
   \\
    Department of Development and Regeneration \\
    KU, Leuven \\
  Belgium \\
  \texttt{} \\
   \And
    Maarten van Smeden
   \\
    Julius Center for Health Sciences and Primary Care \\
    University Medical Center Utrecht \\
  Netherlands \\
  \texttt{} \\
  }

\usepackage{longtable}
\usepackage{float}
\usepackage{threeparttable}
\usepackage{threeparttablex}
\usepackage{pdflscape}
\usepackage{amsmath}
\begin{document}
\maketitle

\begin{abstract}
Risk prediction models are increasingly used in healthcare to aid in
clinical decision making. In most clinical contexts, model calibration
(i.e., assessing the reliability of risk estimates) is critical. Data
available for model development are often not perfectly balanced with
respect to the modeled outcome (i.e., individuals with vs.~without the
event of interest are not equally represented in the data). It is common
for researchers to correct this class imbalance, yet, the effect of such
imbalance corrections on the calibration of machine learning models is
largely unknown. We studied the effect of imbalance corrections on model
calibration for a variety of machine learning algorithms. Using
extensive Monte Carlo simulations we compared the out-of-sample
predictive performance of models developed with an imbalance correction
to those developed without a correction for class imbalance across
different data-generating scenarios (varying sample size, the number of
predictors and event fraction). Our findings were illustrated in a case
study using MIMIC-III data. In all simulation scenarios, prediction
models developed without a correction for class imbalance consistently
had equal or better calibration performance than prediction models
developed with a correction for class imbalance. The miscalibration
introduced by correcting for class imbalance was characterized by an
over-estimation of risk and was not always able to be corrected with
re-calibration. Correcting for class imbalance is not always necessary
and may even be harmful for clinical prediction models which aim to
produce reliable risk estimates on an individual basis.
\end{abstract}


\hypertarget{introduction}{%
\section{Introduction}\label{introduction}}

Risk prediction models are increasingly used in healthcare to aid in
clinical decision making; for example, to help decide if a patient is a
good candidate for surgery or to communicate a patient's risk of disease
\cite{ewout_intro,annals,achilles}. As such, the purpose of a clinical
prediction model is often to estimate a patient's risk of experiencing a
particular event (e.g., successful surgery, disease)
\cite{van_smeden_clinical_2021, moons_prognosis_2009}. Due to the rarity
of many diseases, data available to train clinical prediction models
often exhibit class imbalance i.e., observations from patients with
vs.~without the event of interest are not equally represented in the
data. In machine learning literature, imbalance correction methods are
commonly applied to correct class imbalance by artificially creating
data that are more or perfectly balanced
\cite{cip, summary_m, lp, summary_h}, although the benefit of such
corrections for model performance is not always clear.\\
\strut \\
An abundance of imbalance correction methods exist
\cite{summary_m, lp, summary_h}, yet, information regarding the effect
of these imbalance corrections on model calibration is sparse. Model
calibration captures the accuracy of risk estimates, relating to the
agreement between the estimated (predicted) and observed number of
events \cite{achilles}. In clinical applications where a patient's
predicted risk is the entity used to inform clinical decisions, it is
essential to assess model calibration. If a model is poorly calibrated,
it may produce risk estimates that do not approximate a patient's true
risk well \cite{achilles}. A poorly calibrated model may produce
predicted risks that consistently over- or under-estimate true risk or
that are too extreme (too close to 0 or 1) or too modest (too close to
event prevalence) \cite{achilles}. This can lead to poor treatment
decisions or to clinicians communicating false assurances to patients
\cite{achilles, 12days, bens_paper}. If a clinician uses a poorly
calibrated model to make a highly impactful decision (e.g., to determine
if a patient should receive a bed in an intensive care unit), the costs
of miscalibration to patients are real and can be far-reaching.\\
\strut \\
Some studies have evaluated the effect of class imbalance corrections on
estimation of absolute risks/probabilities and thus, on model
calibration \cite{biesheuvel_advantages_2008, ruben}. Research from van
den Goorbergh and colleagues has demonstrated that class imbalance
corrections may do more harm than good \cite{ruben}; implementing
imbalance corrections resulted in dramatically deteriorated model
calibration, to the point that no corrections were recommended
\cite{ruben}. In the previous research, prediction models were developed
using logistic regression and penalized logistic regression
\cite{biesheuvel_advantages_2008, ruben} . In practice, prediction
models developed for clinical applications increasingly use more
flexible machine learning methods \cite{constanza}. A recent systematic
review of clinical prediction models indicates that machine learning
algorithms like support vector machine and tree-based learning with
random forest, are especially common \cite{constanza}. The impact of
imbalance corrections on model calibration is currently unknown for
prediction models developed using these more flexible machine learning
algorithms.\\
\strut \\
Building on previous research \cite{biesheuvel_advantages_2008, ruben},
we assessed the impact of imbalance corrections on model calibration for
prediction models trained with a wide variety of machine learning
algorithms including: logistic regression, support vector machine,
random forest, XGBoost, RUSBoost and EasyEnsemble. This paper is
structured as follows. We present the design of our simulation study in
section 2 and the results in section 3. A case study where we study the
impact of imbalance corrections using empirical data is presented in
section 4, followed by a discussion of our findings and their
implications for researchers developing prediction models in the
presence of class imbalance in section 5. Our conclusions are presented
in section 6.

\hypertarget{methods}{%
\section{Methods}\label{methods}}

We implemented a simulation study to investigate the effects of
imbalance correction methods across 18 unique data-generating scenarios.
In our research, we focused on prediction models designed for
dichotomous risk prediction. For each scenario, we compared the
out-of-sample predictive performance (i.e., model performance on data
not used to train the model) of models developed with an imbalance
correction to those developed without a correction for class imbalance.
In the next sections, details regarding the data-generating mechanism,
model development procedure, simulation methods, performance measures,
and software and error handling are presented. All code used for this
project is publicly available on GitHub:
\url{https://github.com/alexcarriero/class_imbalance_project}.

\hypertarget{data-generating-scenarios}{%
\subsection{Data Generating Scenarios}\label{data-generating-scenarios}}

In our simulation study, 18 (\(3\) x \(2\) x \(3\)) unique
data-generating scenarios were studied (Table \ref{tab:sim_sets}). This
was achieved by varying the following three characteristics of the data:
the event fraction (proportion of patients with an event), the number of
predictors and sample size. Event fraction varied through the set \{0.5,
0.2, 0.02\} and number of predictors through the set \{8, 16\}. A class
balanced scenario (event fraction = 0.5) was included to study the
effects of imbalance corrections when they are tasked with correcting
for chance imbalances in the simulated data. In all scenarios, data were
generated to yield an expected concordance statistic of \(0.85\). Given
the number of predictors, the event fraction and expected concordance
statistic, we computed the minimum required sample size for a prediction
model (N) developed under these conditions. Sample size calculations
were carried out using the R package pmsampsize \cite{pmsampsize}. The
sample size of data used to train the prediction models
(\(\mathrm{n}_{\mathrm{train}}\)) was then varied through the set
\{\(\frac{1}{2}\)N, N, \(2\)N\} implying half the required sample size,
exactly the required sample size and double the required sample size,
respectively.

\hypertarget{data-generating-mechanism}{%
\subsection{Data Generating Mechanism}\label{data-generating-mechanism}}

As we focused on dichotomous risk prediction, we generated data
comprised of two classes. We refer to the non-events and events as class
0 and class 1, respectively. Data for each class were generated
independently using distinct multivariate normal (\(mvn\))
distributions. As shown in equations (1) and (2), we specified a
distinct mean and covariance structure for each class. The mean
structures for the classes are represented as \(\mu_0\) and \(\mu_1\)
for class 0 and 1, respectively. The covariance matrices are represented
as \(\Sigma_0\) and \(\Sigma_1\) for class 0 and 1, respectively.

\begin{align}
&\mathrm{Class \ \  0:} \mathbf{X} \sim mvn( \pmb{\mu_0}, \pmb{\Sigma_0}) = mvn(\pmb{0}, \pmb{\Sigma_0}), \\
&\mathrm{Class \ \  1:} \mathbf{X} \sim mvn( \pmb{\mu_1}, \pmb{\Sigma_1}) = mvn(\pmb{\Delta_\mu}, \pmb{\Sigma_0} - \pmb{\Delta_\Sigma}).
\end{align}\\

As shown in equation (2), the differences in parameter values between
the two classes are represented by \(\Delta_\mu\) and \(\Delta_\Sigma\);
a vector and matrix comprised of the differences in predictor means, and
variances/ covariances, between the classes, respectively. We specified
no variation in means among predictors within a class, making all
elements in \(\Delta_\mu\) equivalent; we denote these equivalent
elements as \(\delta_\mu\). Similarly, we specified no variation in
predictor variances within a class, making all diagonal elements in the
matrix \(\Delta_\Sigma\) equivalent, denoted by \(\delta_\Sigma\).\\
\strut \\
For class 0, all predictor means were fixed to zero and all variances to
1. Consequently for class 1, all predictor means were equal to
\(\delta_\mu\) and all variances equal to 1 - \(\delta_\Sigma\).
Finally, in both classes, we allowed \(75\)\% of the predictors to
covary. All non-zero correlations among predictors in each class were
set to \(0.2\). To ensure the correlation among predictors was not
stronger in one class, we fixed the correlation matrices of the two
classes to be equal. This was accomplished by computing the off-diagonal
elements of \(\Sigma_1\) such that the correlation matrices of the two
classes were equivalent (as shown below).\\
\strut \\
For instance, with 8 predictors, the mean and covariance structure for
class 0 was,\\

\begin{equation*}
\pmb{\mu_0} = \begin{bmatrix}
 0 \\ 0 \\ 0\\ 0 \\ 0 \\ 0 \\ 0 \\ 0
\end{bmatrix}, \pmb{\Sigma_0} = \begin{bmatrix}
1   & 0.2 & 0.2 & 0.2 & 0.2 & 0.2 & 0 & 0\\
0.2 & 1   & 0.2 & 0.2 & 0.2 & 0.2 & 0 & 0\\
0.2 & 0.2 & 1   & 0.2 & 0.2 & 0.2 & 0 & 0\\
0.2 & 0.2 & 0.2   & 1 & 0.2 & 0.2 & 0 & 0\\
0.2 & 0.2 & 0.2   & 0.2 & 1 & 0.2 & 0 & 0\\
0.2 & 0.2 & 0.2   & 0.2 & 0.2 & 1 & 0 & 0\\
0   & 0   & 0     &  0 & 0    & 0 & 1 & 0\\
0   & 0   & 0     &  0 & 0    & 0 & 0 & 1\\
\end{bmatrix},
\end{equation*}

\hfill\break
\hfill\break
and mean and covariance structure for class 1 was,\\

\begin{equation*}
\pmb{\mu_1} = \begin{bmatrix}
 \delta_\mu \\ \delta_\mu \\ \delta_\mu \\ \delta_\mu \\ \delta_\mu \\ \delta_\mu \\ \delta_\mu \\ \delta_\mu
\end{bmatrix}, \pmb{\Sigma_1} = \begin{bmatrix}
1 - \delta_\Sigma   & z & z & z & z & z & 0 & 0\\
z & 1 - \delta_\Sigma   & z & z & z & z & 0 & 0\\
z & z & 1 - \delta_\Sigma   & z & z & z & 0 & 0\\
z & z & z   & 1 - \delta_\Sigma & z & z & 0 & 0\\
z & z & z   & z & 1 - \delta_\Sigma & z & 0 & 0\\
z & z & z   & z & z & 1 - \delta_\Sigma & 0 & 0\\
0   & 0   & 0     &  0 & 0    & 0 & 1 - \delta_\Sigma & 0\\
0   & 0   & 0     &  0 & 0    & 0 & 0 & 1 - \delta_\Sigma\\
\end{bmatrix}.
\end{equation*}\\
\strut \\
Here, \(z = (1-\delta_\Sigma)*0.2\), to ensure equivalent correlation
matrices between the two classes.\\
\strut \\
Parameter values for the data generating distributions (\(\delta_\mu\)
and \(\delta_\Sigma\)) of each scenario were selected to generate a
concordance statistic (\(C\)) of \(0.85\). Under the assumption of
normality for all predictors (in each class), the concordance statistic
of the data can be expressed as a function of \(\Delta_\mu\),
\(\Sigma_0\) and \(\Sigma_1\) \cite{mvauc}. Optimal values of
\(\delta_\mu\) and \(\delta_\Sigma\) for each scenario were computed
analytically, based on the following formula \cite{mvauc}:\\
\begin{equation}
C = \Phi \left( \sqrt{\pmb{\Delta_\mu}{'}\  (\pmb{\Sigma_0} + \pmb{\Sigma_1})^{-1} \ \pmb{\Delta_{\mu}}} \right).
\end{equation}\\
In equation (3), \(\Phi\) represents the cumulative density function of
the standard normal distribution; \(\Delta_\mu\), \(\Sigma_0\) and
\(\Sigma_1\) maintain their previous definitions. To ensure a unique
solution, \(\delta_\Sigma\) was fixed at 0.3 for each scenario, while
equation (3) was solved to yield the appropriate value of \(\delta_\mu\)
in each scenario. The full analytical solution is provided in Appendix
A. The parameter values for the data generating distributions in each
simulation scenario are presented in Table \ref{tab:sim_sets}.\\
\strut \\
Given that data for each class were generated independently, we had
direct control over how many observations were generated under each
class. The number of events (\(n_1\)) was sampled from the binomial
distribution with probability equal to the specified event fraction. The
number of non-events (\(n_0\)) was then computed as \(n - n_1\), where
\(n\) is the specified sample size for a given scenario.

\hypertarget{model-development}{%
\subsection{Model Development}\label{model-development}}

All prediction models were developed according to the following two-step
procedure. First, data were pre-processed using a class imbalance
correction technique. Then, the resulting artificially-balanced data
were used to train a machine learning algorithm.\\
\strut \\
We implemented a \(5\) x \(6\) full-factorial design to compare the
out-of-sample predictive performance of prediction models developed with
\(5\) imbalance corrections (\(1\) control and \(4\) corrections) and
\(6\) machine learning algorithms. As a control, data were not corrected
for imbalance and the uncorrected data went on to train the machine
learning algorithms. In total, we compared the performance of 30
prediction models, each comprised of a unique combination of imbalance
correction and machine learning algorithm.

\hypertarget{imbalance-corrections}{%
\subsubsection{Imbalance Corrections}\label{imbalance-corrections}}

\begin{figure}

{\centering \includegraphics[width=0.9\linewidth]{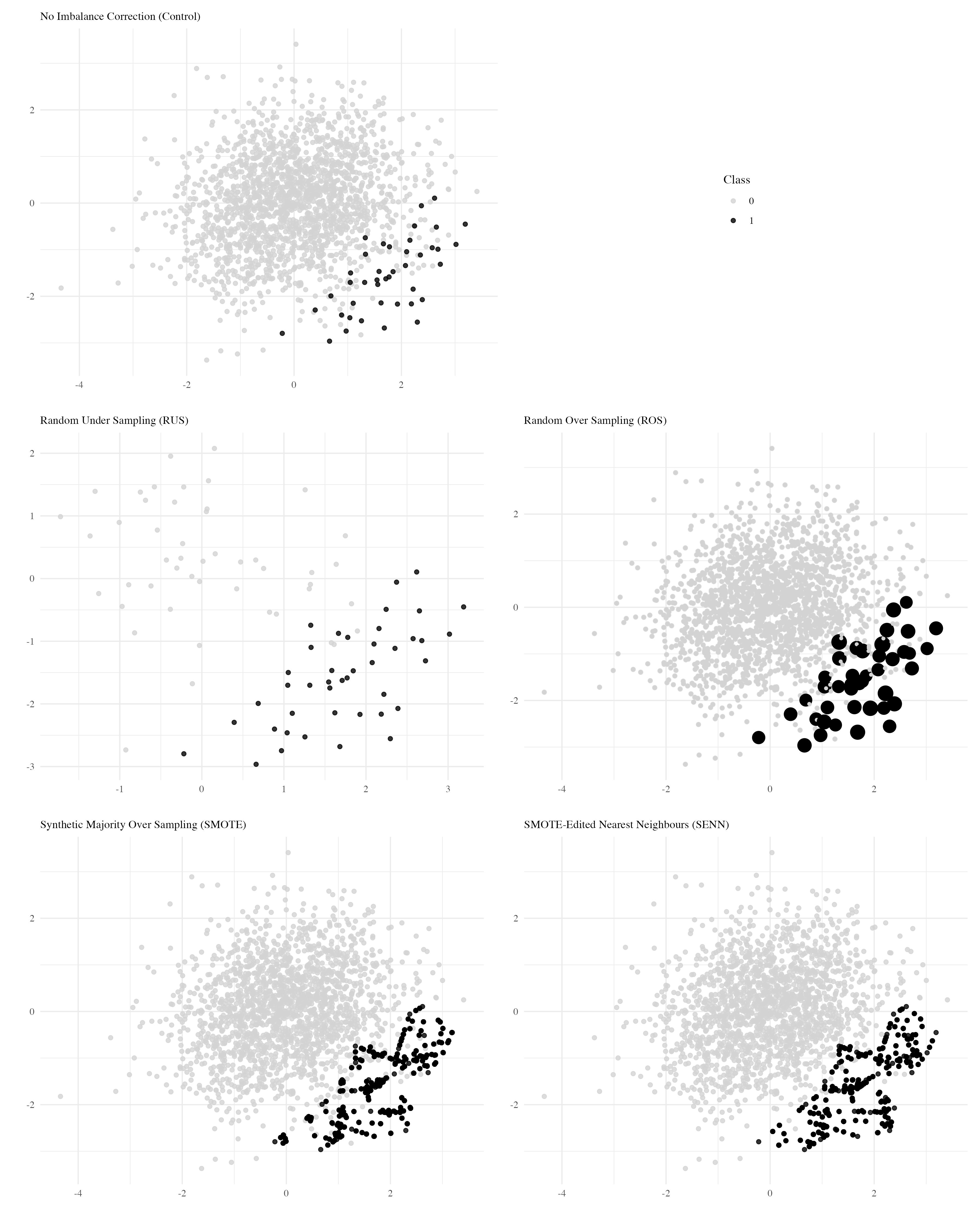} 

}

\caption{An illustration of the data pre-processing  imbalance corrections studied in our simulation.  The data shown have a sample size of 2000 and an event fraction of 0.02, imbalance corrections were applied to achieve a target event fraction of 0.5 (perfect balance). \label{fig:imb_corr}}\label{fig:unnamed-chunk-1}
\end{figure}

The imbalance corrections studied in our simulation included four data
pre-processing techniques: random under sampling (RUS), random over
sampling (ROS), synthetic minority over sampling (SMOTE), synthetic
minority over sampling with Wilson's Edited Nearest Neighbor Rule
(SENN), and a control in which no correction was implemented. As
determined by a recent systematic review, RUS, ROS and SMOTE are the
imbalance corrections that are commonly implemented when developing
clinical prediction models, SMOTE being the most popular among them
\cite{constanza}. We included SENN as well, given literature indicating
that it can outperform SMOTE under certain conditions
\cite{senn, heart_failure_senn}. All imbalance corrections and the R
packages used for their implementation in our simulation study are
summarized in Table \ref{tab:corrections}.\\
\strut \\
The functionality of these data pre-processing techniques is illustrated
in Figure \ref{fig:imb_corr}. In our simulation, imbalance corrections
were implemented to achieve artificially class balanced data (event
fraction \(\approx 0.5\)). RUS achieves class balance by randomly
disregarding observations from the majority class until a balance is
achieved. This technique is similar to a case-control design without
matching \cite{rutjes_casecontrol_2005}. ROS achieves balance by
randomly re-sampling from the minority class with replacement until a
balance is achieved. SMOTE generates artificial observations for the
minority class by interpolating from the existing minority class
observations \cite{chawla}. In our implementation, we specified the
number of nearest neighbors for SMOTE (\(k\)) to be 5 (package default)
\cite{iric}. In SENN, class balance is achieved by using a combination
of SMOTE and Wilson's Edited Nearest Neighbor Rule (ENN) \cite{wilson}.
SMOTE is implemented first, to generate synthetic data for the minority
class and then ENN is implemented to remove any observation which has a
different outcome than the majority of its nearest neighbors
\cite{senn}. This added step is implemented to make it easier for a
classifier to distinguish between the classes, as when synthetic
observations are generated for the minority class using SMOTE, it may
cause an increase in noise near the class boundary. In our
implementation, we specified the number of nearest neighbors in the
SMOTE step (\(k_1\)) to be 5, and the number of nearest neighbors in ENN
step (\(k_2\)) to be 3 (package defaults) \cite{iric}. Finally, in the
control condition, data were not corrected for imbalance and moved
directly to the second step of model development untouched (i.e., the
imbalanced data were used to train the prediction models).

\hypertarget{machine-learning-algorithms}{%
\subsubsection{Machine Learning
Algorithms}\label{machine-learning-algorithms}}

Machine learning algorithms were selected based on a recent systematic
review identifying common algorithms used to develop prediction models
in a medical context \cite{constanza}. These algorithms include:
logistic regression (LR), support vector machine (SVM), random forest
(RF) and XGBoost (XG). Additionally, based on literature summarizing
common strategies to handle class imbalance \cite{summary_m, lp, kaur},
we included two ensemble learning algorithms designed specifically to
handle class imbalance: RUSBoost (RB) and EasyEnsemble (EE). While both
of these algorithms use random undersampling innately, RUSBoost
\cite{rusboost} is a boosting algorithm while EasyEnsemble \cite{ee}
uses bagging. The hyperparameters selected for these machine learning
algorithms, and the R packages used for algorithm implementation are
summarized in Table \ref{tab:algs}.\\
\strut \\
Four machine learning algorithms (SVM, RF, XG and RB) required the
specification of model hyperparameters. Hyperparameter tuning was
conducted using the R package caret for SVM, RF and XG \cite{caret}. The
methods of implementation were svmRadial, ranger and xgbTree for SVM, RF
and XG, respectively. To select the hyperparameters, we implemented
5-fold cross-validation optimizing for model deviance. For SVM and XG,
caret default tune grids and tune length were used. For RF, we specified
a custom tuning grid: mtry (the number of candidate splitting variables
allowed at each node in a tree) was allowed to vary from 1 to the total
number of predictors, min.node.size (the minimum number of observations
allowed in a leaf node) was allowed to vary from 1 to 10 and we
specified `gini' as the splitrule (the split which minimizes Gini
impurity). Finally, in our implementation of RUSBoost we specified a
support vector machine with a radial kernel as the weak classifier and
an ensemble size of 10 (package defaults) \cite{ebmc}.\\
\strut \\
Notably, since RB and EE innately correct for class imbalance, there are
only 4/30 prediction models for which no attempt to correct for
imbalance is made. Given our 5x6 full-factorial design, we are able to
compare the performance of the models which make no attempt to correct
for class imbalance (control models for LR, SVM, RF, XG) with those
which correct for class imbalance via a data pre-processing technique
(RUS, ROS, SMOTE, SENN), innately correct for class imbalance (RB, EE),
or both.

\hypertarget{simulation-methods}{%
\subsection{Simulation Methods}\label{simulation-methods}}

Under each simulation scenario, \(2000\) data sets were generated. Each
data set was comprised of training and validation data. The training and
validation data were generated independently using identical data
generating mechanisms. Validation data sets were generated to be ten
times larger than the training data sets. The sample sizes of the
training (\(\mathrm{n}_{\mathrm{train}}\)) and validation
(\(\mathrm{n}_{\mathrm{validation}}\)) data for each simulation scenario
can be found in Table \ref{tab:sim_sets}.\\
\strut \\
For each generated data set, 30 (\(5\) x \(6\)) prediction models were
developed (as described in section 2.3). All prediction models were
trained using the training data. Predictive performance was then
assessed using the validation data.\\
\strut \\
Since we expected many machine learning algorithms to exhibit
miscalibration, especially in scenarios with considerable class
imbalance, we also conducted logistic (re-)calibration for all
prediction models, using the following procedure. For each observation
in the validation set, the predicted risk and corresponding observed
outcome (0 or 1) were stored. Then, the predicted risks were
re-calibrated using the following logistic regression model:\\
\strut \\
\begin{equation}
\mathrm{log} \left( \frac{P(Y_i = 1)}{1- P(Y_i = 1)} \right) = \beta_0 + \mathrm{log} \left( \frac{p_i}{1-p_i} \right).
\end{equation}\\
\strut \\
Here, \(Y_i\) represents the observed outcome for the \(i\)th
observation in the validation set and \(p_i\) represents the predicted
risk for the \(i\)th observation in the validation set, from a given
prediction model. The logarithms in equation (4) represent the natural
logarithm. This approach is similar to platt scaling \cite{platt},
except, only an intercept term is estimated (\(\beta_0\)), while platt
scaling typically includes the estimation of both an intercept and slope
(i.e., a slope coefficient is estimated, rather than included as an
offset) \cite{platt}. After the re-calibration procedure was
implemented, predictive performance was then re-assessed using the
re-calibrated predictions.\\
\strut \\
In summary, predictive performance was assessed twice in our simulation:
first using the raw predictions (no re-calibration) and subsequently
using the re-calibrated predictions.

\hypertarget{performance-measures}{%
\subsection{Performance Measures}\label{performance-measures}}

Out-of-sample predictive performance was assessed using measures of
calibration, discrimination and overall performance. All performance
measures were computed using the raw predicted risks (resulting from the
validation data) and subsequently, using the re-calibrated predictions,
for each prediction model.\\
\strut \\
Model calibration measures the agreement between predicted risks and
observed proportions of the event in the data \cite{achilles}.
Calibration was assessed using both visual and empirical metrics. We
assessed calibration visually by means of flexible calibration curves;
one flexible calibration curve was generated for each simulation
iteration. Coordinates for the calibration curves were calculated using
loess regression; implemented using the R package stats \cite{r}.
Calibration curves were then generated using ggplot2 \cite{gg}.
Additionally, calibration intercept and slope were calculated according
to their respective definitions in Steyerberg \emph{et al}. (2010)
\cite{epi}. In a flexible calibration curve, when predicted risks
(x-axis) correspond well with the observed proportions in the data
(y-axis), the curve follows a straight diagonal line (\(y = x\))
\cite{achilles}. With respect to calibration intercept and slope, ideal
calibration is represented by values of 0 and 1, respectively
\cite{epi}.\\
\strut \\
The concordance statistic (C) was used to measure model discrimination;
computed using the R package pROC \cite{pROC}. This metric captures a
model's ability to yield higher risk estimates for patients with the
event than for those without the event. For dichotomous outcomes, it is
equivalent to the area under the Receiver Operator Characteristic curve
\cite{epi}. A model which perfectly discriminates between the classes
will have C = 1; a model with no discriminative performance has C = 0.5
\cite{epi}.\\
\strut \\
Overall performance was measured by Brier score. This metric reflects
both model discrimination and calibration and was calculated according
to its definition in Steyerberg \emph{et al}. (2010) \cite{epi}. In an
ideal model, predicted risks approximate the observed outcome well for
all individuals; perfect prediction models produce a Brier score of
zero. As event fraction decreases, it is easier for a prediction model
to achieve a low Brier score \cite{epi}. Comparisons for this
performance measure should only be made among prediction models
developed with the same event fraction.\\
\strut \\
For empirical measures of model performance (concordance statistic,
Brier score, calibration intercept and calibration slope), the median
over the simulation iterations and corresponding Monte Carlo error were
reported. The values of each performance measure across all 2000
simulation iterations were also visualized with violin plots, generated
with ggplot2 \cite{gg}.

\hypertarget{software-and-error-handling}{%
\subsection{Software and Error
Handling}\label{software-and-error-handling}}

The simulation study was conducted using the University Medical Center
Utrecht's high performance computing (HPC). This high performance
computer uses two types of central processing units: Intel(R) Xeon (R)
Silver and Intel(R) Xeon (R) Gold. The simulation study and processing
of results were conducted using R versions 4.2.2 and 4.1.2, respectively
\cite{r}. For further details (e.g., R package versions, HPC
specifications) please see Supplementary Materials (Section A).\\
\strut \\
Any warnings or errors which occurred during the simulation study were
carefully monitored and saved. In case of an error with an imbalance
correction or a machine learning algorithm, no new data were generated.
If an error was produced by an imbalance correction, uncorrected data
were allowed to proceed in the model development process (i.e., the
uncorrected data were used to train the machine learning algorithms). In
case of an error produced by a machine learning algorithm, predicted
risks were saved as missing (NA). Consequently, empirical performance
metrics were stored as missing and no flexible calibration curve was
generated. Please see Supplementary Materials (Section B) for reporting
of all warnings or errors generated.

\hypertarget{results}{%
\section{Results}\label{results}}

In the next sections, we focus on the results for simulation scenarios
with 8 predictors, sample size equivalent to the minimum required sample
sample size (N) and balanced (event fraction \(= 0.5\)), moderately
imbalanced (event fraction \(=0.2\)) and strongly imbalanced (event
fraction \(=0.02\)) data, simulation scenarios 4-6, respectively.
Results from all simulation scenarios are included in the Supplementary
Materials (Section C) and are also presented in our Shiny App:
\url{https://alex-carriero.shinyapps.io/class_imbalance/}.\\
\strut \\
We present the results from scenarios 4-6 because they are
representative of the results across all 18 data-generating scenarios.
Results did not vary greatly across the number of predictors or sample
size settings considered (Appendix B). In particular, increasing the
number of predictors or sample size slightly improved model
discrimination for all prediction models, meanwhile, calibration
intercept, calibration slope and brier score remained unchanged (Figures
B1-B4, Appendix B).

\hypertarget{calibration}{%
\subsection{Calibration}\label{calibration}}

\begin{figure}

{\centering \includegraphics[width=0.9\linewidth]{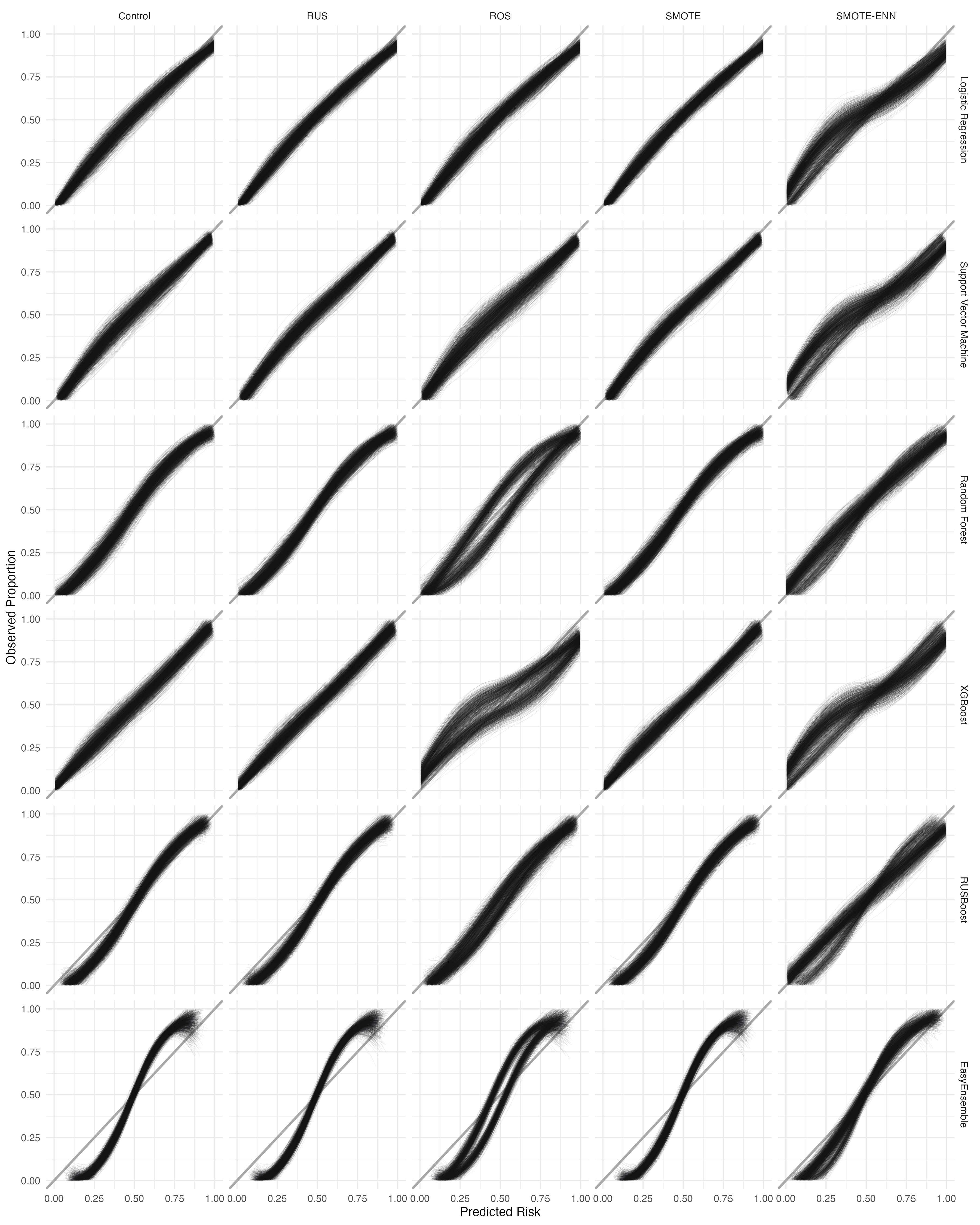} 

}

\caption{Flexible calibration curves for each of 2000 simulation iterations in simulation scenario 4. This simulation scenario is characterized by 8 predictors, exactly the minimum required sample size (N) and class balanced data (event fraction = 0.5). Flexible curves were generated using raw predicted risks; no re-calibration. Imbalance corrections: RUS (random undersampling), ROS (random oversampling), SMOTE (synthetic minority oversampling), SMOTE-ENN (synthetic minority oversampling with Wilson's Edited Nearest Neighbor rule). \label{fig:plot1}}\label{fig:unnamed-chunk-2}
\end{figure}

\begin{figure}

{\centering \includegraphics[width=0.9\linewidth]{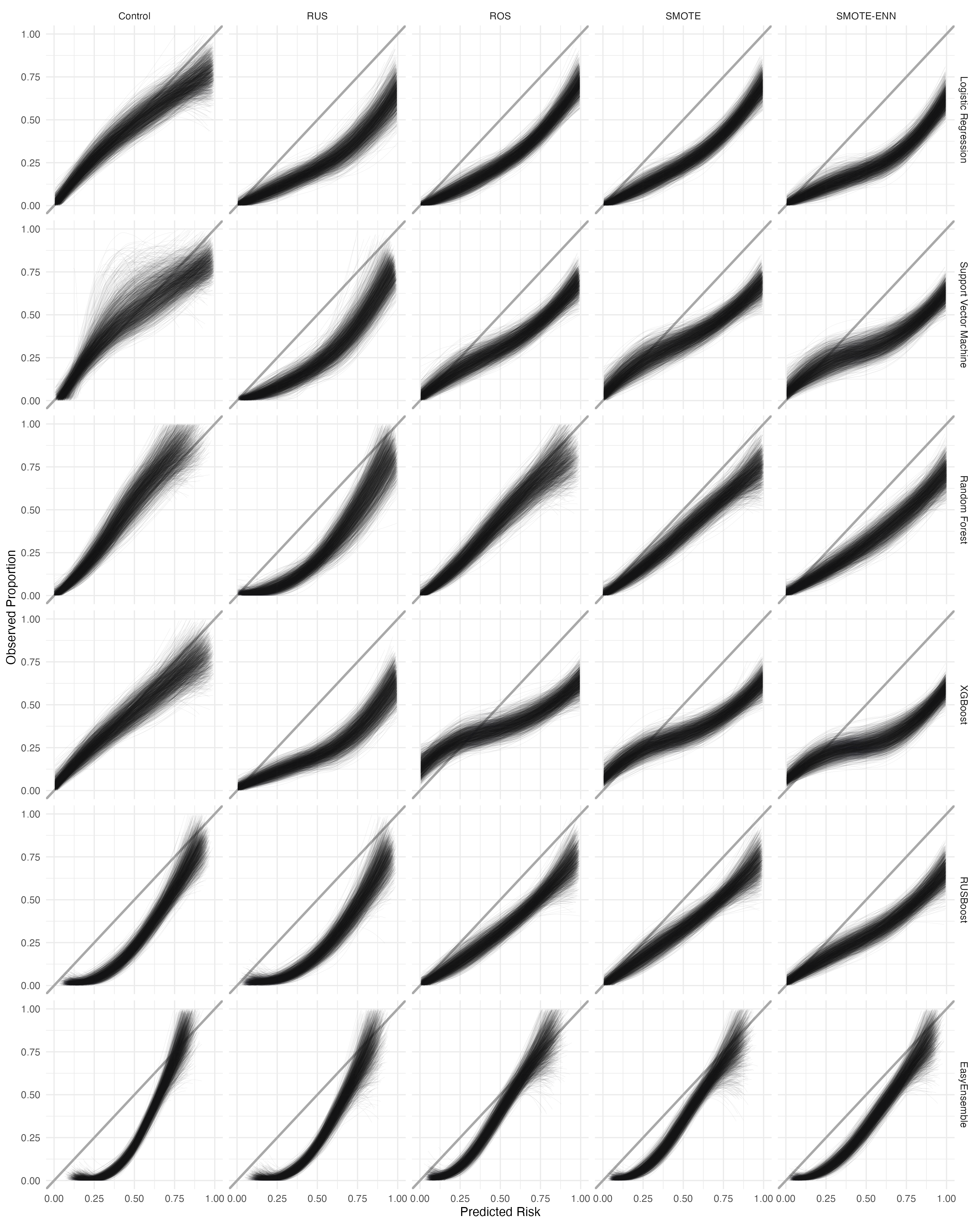} 

}

\caption{Flexible calibration curves for each of 2000 simulation iterations in simulation scenario 5. This simulation scenario is characterized by 8 predictors, exactly the minimum required sample size (N) and moderately imbalanced data (event fraction = 0.2). Flexible curves were generated using raw predicted risks; no re-calibration. Imbalance corrections: RUS (random undersampling), ROS (random oversampling), SMOTE (synthetic minority oversampling), SMOTE-ENN (synthetic minority oversampling with Wilson's Edited Nearest Neighbor rule). \label{fig:plot2}}\label{fig:unnamed-chunk-3}
\end{figure}

\begin{figure}

{\centering \includegraphics[width=0.9\linewidth]{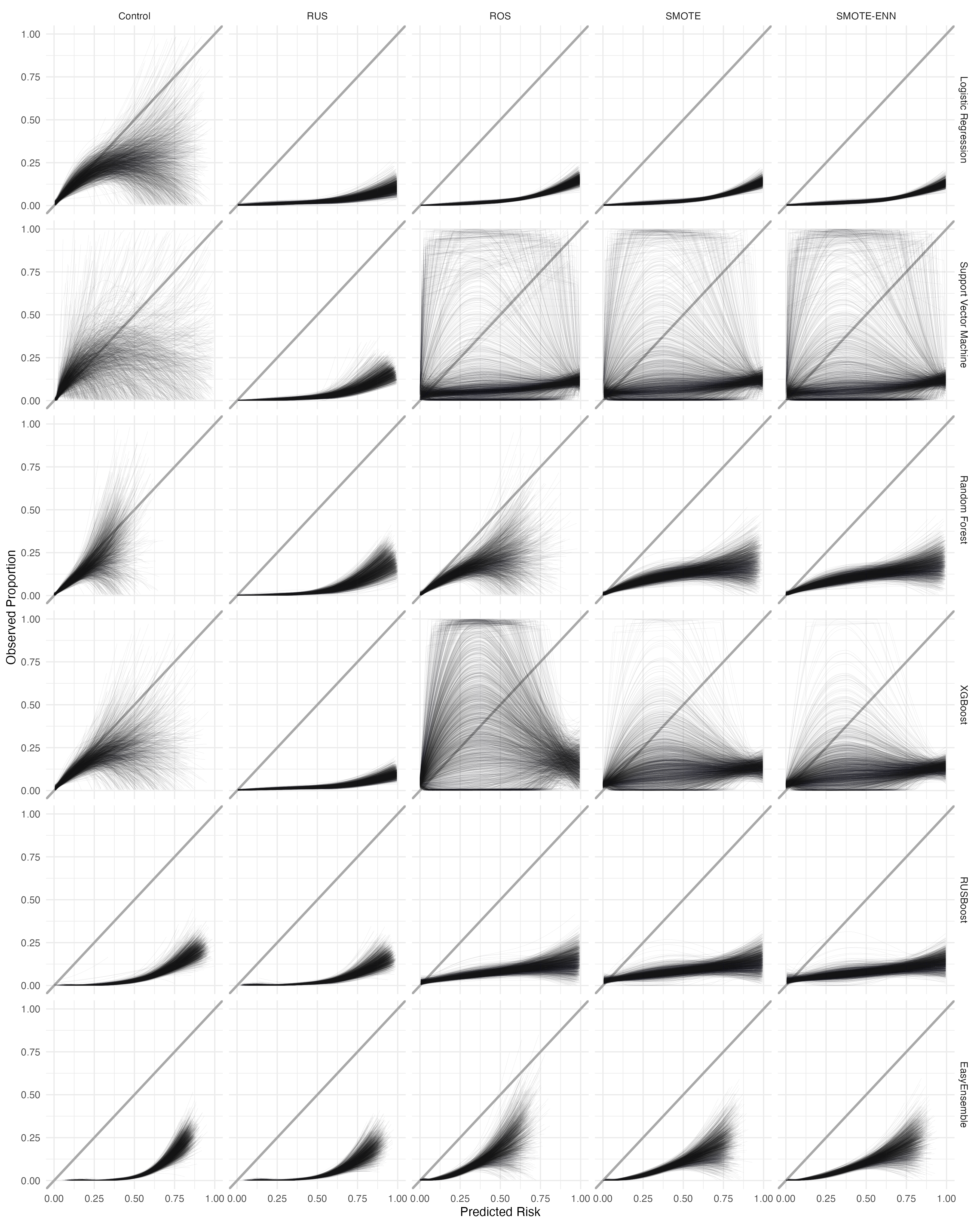} 

}

\caption{Flexible calibration curves for each of 2000 simulation iterations in simulation scenario 6. This simulation scenario is characterized by 8 predictors, exactly the minimum required sample size (N) and strongly imbalanced data (event fraction = 0.02). Flexible curves were generated using raw predicted risks; no re-calibration. Imbalance corrections: RUS (random undersampling), ROS (random oversampling), SMOTE (synthetic minority oversampling), SMOTE-ENN (synthetic minority oversampling with Wilson's Edited Nearest Neighbor rule). \label{fig:plot3}}\label{fig:unnamed-chunk-4}
\end{figure}

\begin{figure}

{\centering \includegraphics[width=0.9\linewidth]{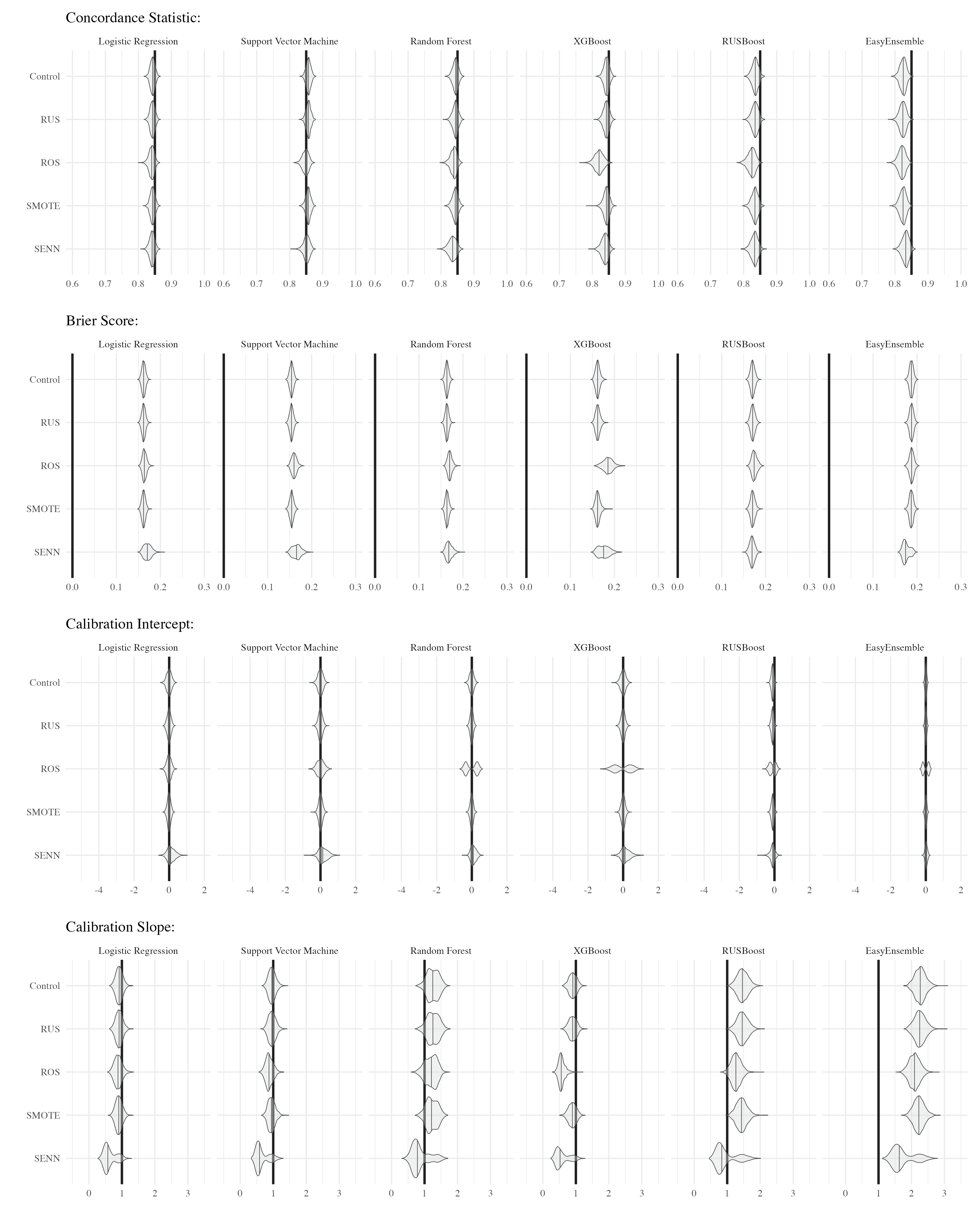} 

}

\caption{Empirical performance metrics for each of 2000 simulation iterations in simulation scenario 4.  This simulation scenario is characterized by 8 predictors, exactly the minimum required sample size (N) and class balanced data (event fraction = 0.5). Empirical performance metrics were computed with raw predicted risks; no re-calibration. The target value highlighted for the concordance statistic reflects the data-generating concordance statstic, 0.85. Imbalance corrections: RUS (random undersampling), ROS (random oversampling), SMOTE (synthetic minority oversampling), SENN (synthetic minority oversampling with Wilson's Edited Nearest Neighbor rule). \label{fig:plot4}}\label{fig:unnamed-chunk-5}
\end{figure}

\begin{figure}

{\centering \includegraphics[width=0.9\linewidth]{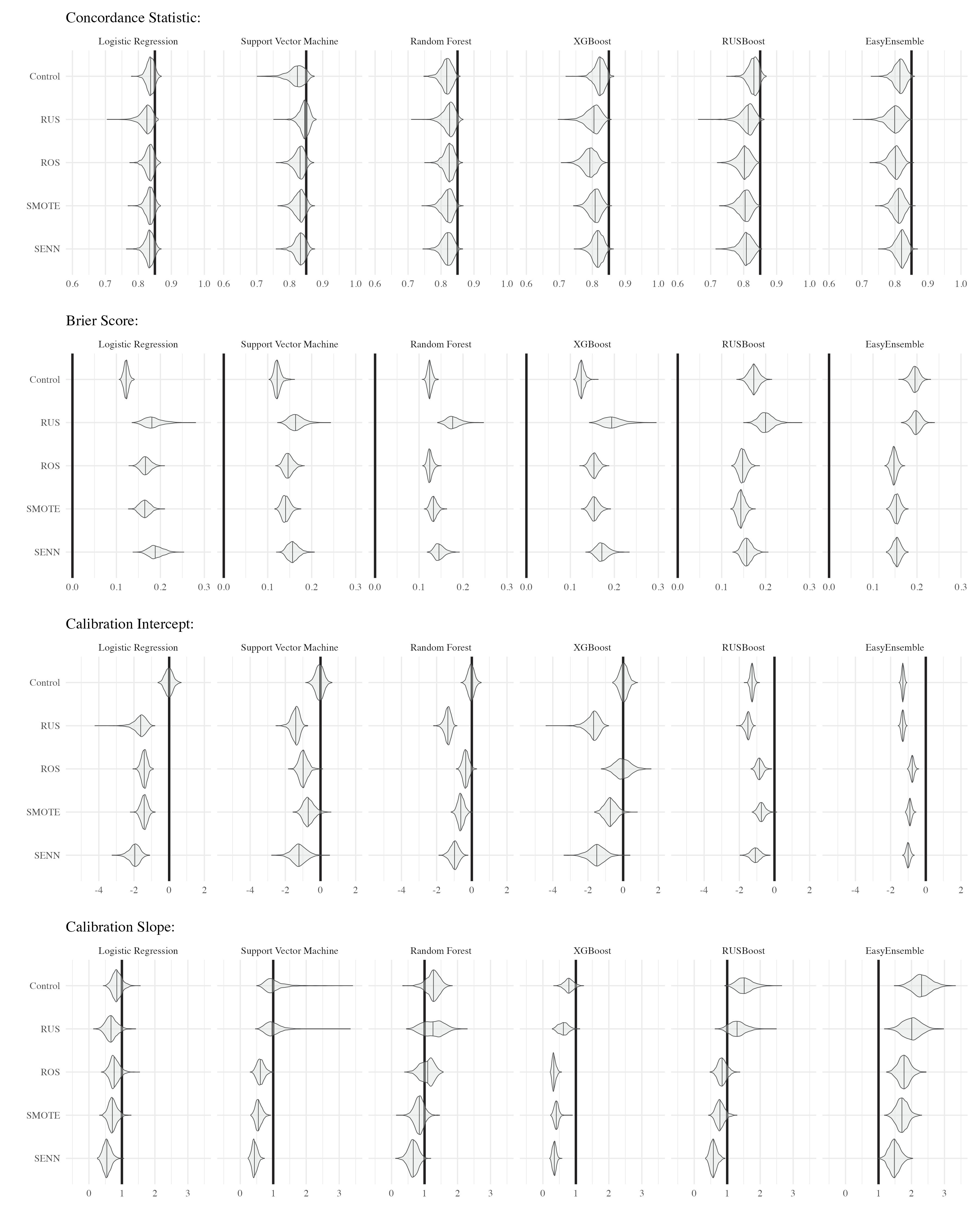} 

}

\caption{Empirical performance metrics for each of 2000 simulation iterations in simulation scenario 5.  This simulation scenario is characterized by 8 predictors, exactly the minimum sample size (N) and moderately imbalanced data (event fraction = 0.2). Empirical performance metrics were computed with raw predicted risks; no re-calibration. The target value highlighted for the concordance statistic reflects the data-generating concordance statstic, 0.85. Imbalance corrections: RUS (random undersampling), ROS (random oversampling), SMOTE (synthetic minority oversampling), SENN (synthetic minority oversampling with Wilson's Edited Nearest Neighbor rule). \label{fig:plot5}}\label{fig:unnamed-chunk-6}
\end{figure}

\begin{figure}

{\centering \includegraphics[width=0.9\linewidth]{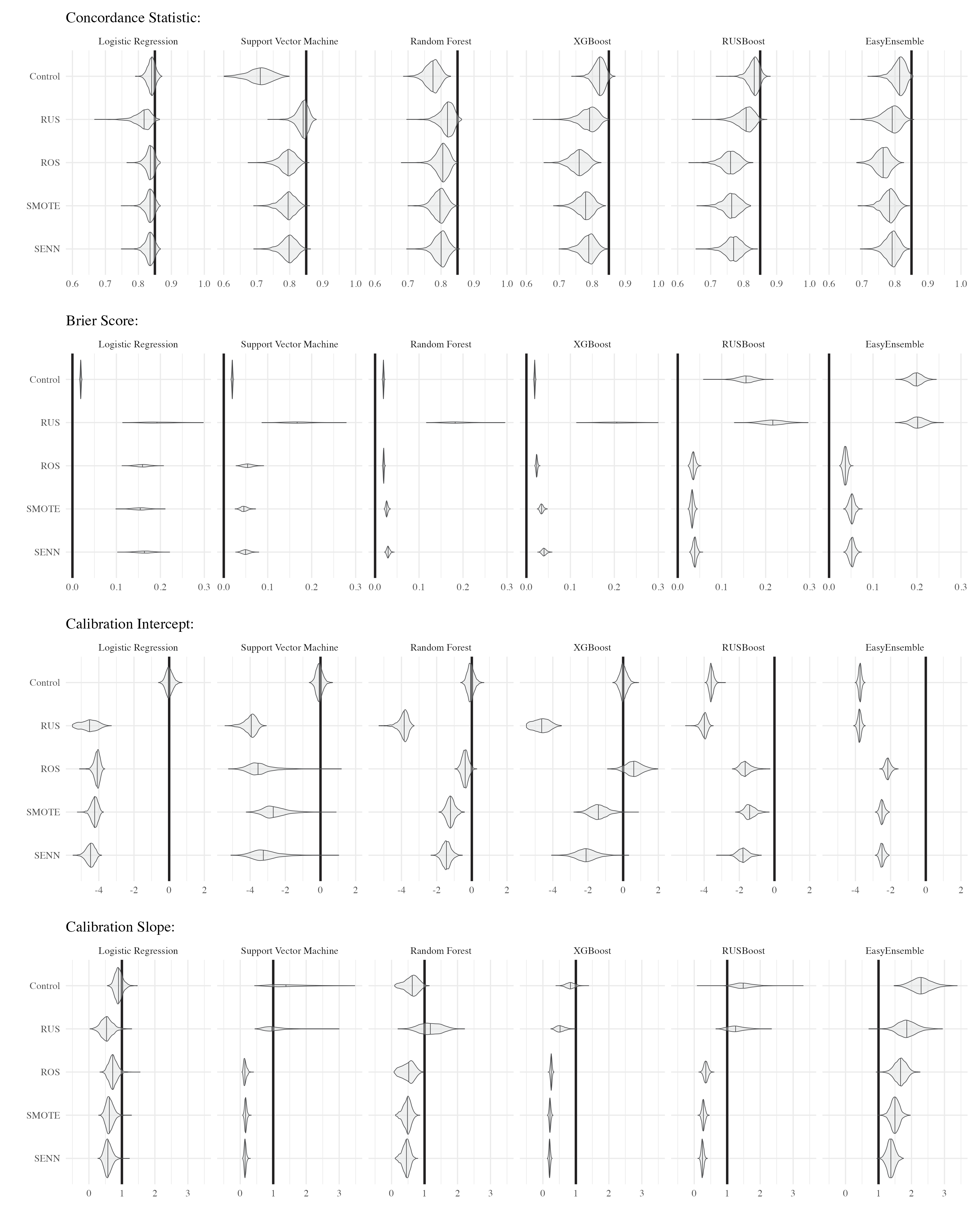} 

}

\caption{Empirical performance metrics for each of 2000 simulation iterations in simulation scenario 6.  This simulation scenario is characterized by 8 predictors, exactly the minimum required sample size (N) and strongly imbalanced data (event fraction = 0.02). Empirical performance metrics were computed with raw predicted risks; no re-calibration. The target value highlighted for the concordance statistic reflects the data-generating concordance statstic, 0.85. Imbalance corrections: RUS (random undersampling), ROS (random oversampling), SMOTE (synthetic minority oversampling), SENN (synthetic minority oversampling with Wilson's Edited Nearest Neighbor rule). \label{fig:plot6}}\label{fig:unnamed-chunk-7}
\end{figure}

For balanced data (event fraction = 0.5) all machine learning
algorithms, except EE, were well calibrated when training data were
pre-processed with the control, RUS or SMOTE (Figure \ref{fig:plot1});
this was reflected by calibration intercepts and slopes very near to 0
and 1, respectively (Figure \ref{fig:plot4}). Interestingly, when
training data were pre-processed with ROS, there was separation between
the calibration curves (Figure \ref{fig:plot1}) for models using
tree-based algorithms (RF, XG, EE). This separation was reflected by
separation in the calibration intercepts (Figure \ref{fig:plot4}). The
division among the flexible calibration curves and calibration
intercepts occurred as a result of chance imbalance in the training data
(approximately half the observed event fractions were \textgreater{} 0.5
while the other half were \textless{} 0.5). The top-curves (calibration
intercepts \textgreater{} 0) underestimated risk on average and were
generated when chance imbalance favored events (i.e., more events than
non-events). Bottom-curves (calibration intercepts \textless{} 0)
overestimated risk on average and were generated when chance imbalance
favored non-events. When SENN was used to pre-process the training data,
LR, SVM and XG exhibited worse calibration than their controls,
meanwhile RF, RB and EE remained as well calibrated (RF and RB) or
slightly better calibrated (EE) than their controls. In our reporting,
for a given machine learning algorithm, the control refers to a model
trained with data that were not corrected for imbalance. With respect to
EE, regardless of the data pre-processing method, the predicted risks
were too moderate (calibration slopes \textgreater{} 1, Figure
\ref{fig:plot4}).\\
\strut \\
When data exhibited moderate imbalance (event fraction = 0.2), control
models for LR, SVM, RF and XG were well calibrated, while control models
for RB and EE were not (Figure \ref{fig:plot2}). In the RB and EE
control models, predicted risks consistently over-estimated true risk
(Figure \ref{fig:plot2}); this miscalibration was characterized by
median calibration intercepts below 0 (RB: -1.27, EE: -1.3) and median
calibration slopes above 1 (RB: 1.52, EE: 2.31), as shown in Figure
\ref{fig:plot5}. Similarly, when training data were pre-processed with
any imbalance correction, prediction models produced risk estimates
which consistently over-estimated true risk, regardless of the machine
learning algorithm used (Figure \ref{fig:plot2}). RF trained with ROS
pre-processed data was protected against this general trend; for this
one prediction model, model calibration was preserved (Figure
\ref{fig:plot2}).\\
\strut \\
When data were strongly imbalanced (event fraction = 0.02), all
prediction models exhibited miscalibration (Figure \ref{fig:plot3}).
With respect to the control models, calibration curves for LR, SVM, RF
and XG were unstable; there was large variation among the calibration
curves produced over the simulation iterations. Meanwhile, the control
models for RB and EE produced predicted risks which exhibited a specific
pattern of miscalibration: all consistently over-estimated true risk,
with very little variation among the calibration curves. Similarly, when
training data were pre-processed with any imbalance correction,
prediction models, again, produced risk estimates which largely
over-estimated true risk. For oversampling corrections (ROS, SMOTE,
SENN) calibration curves for SVM and XG models were extremely unstable.
From Figure \ref{fig:plot6}, it is clear that in this scenario, only the
models which did not attempt to correct for class imbalance (the control
models for LR, SVM, RF and XG) had calibration intercepts centered
around 0. With respect to calibration slope, oversampling corrections
(ROS, SMOTE, SENN) caused a decrease in calibration slope compared to
controls (Figure \ref{fig:plot6}); this resulted in worse calibration
slopes for all algorithms except EE. Meanwhile, the undersampling
correction (RUS) actually improved calibration slopes for SVM and RF
compared to controls.\\
\strut \\
Overall, as imbalance between the classes was magnified, model
calibration deteriorated for all prediction models. All imbalance
corrections affected model calibration in a very similar fashion.
Correcting for imbalance using pre-processing methods (RUS, ROS, SMOTE,
SENN) and/or by using an imbalance correcting algorithm (RB, EE)
resulted in prediction models which consistently over-estimated risk. On
average, no model trained with imbalance corrected data outperformed the
control models in which no imbalance correction was made, with respect
to model calibration.

\hypertarget{discrimination}{%
\subsection{Discrimination}\label{discrimination}}

For class balanced data (event fraction = 0.5), when comparing models
developed with the same machine learning algorithm, all models had equal
or nearly equal discrimination; imbalance corrections had little or no
effect on model discrimination (Figure \ref{fig:plot4}). SVM models had
the highest discrimination: all five models built with SVM (regardless
of the data pre-processing method used) had higher median discrimination
than any other prediction models in this class balanced scenario (Table
\ref{tab:results}). Median concordance statistics in this simulation
scenario ranged from {[}0.82, 0.86{]}.\\
\strut \\
For moderately imbalanced data (event fraction = 0.2), the effects of
the imbalance corrections on model discrimination were highly dependent
on the machine learning algorithm (Figure \ref{fig:plot5}). On average,
RUS and ROS decreased discrimination for LR, XG, RB and EE compared to
controls, meanwhile, on average, they improved discrimination for SVM
and RF compared to controls. SMOTE and SENN had, on average, no effect
or worsened discrimination for LR, RF, XG, RB and EE, while they
improved discrimination for SVM. These effects were small and are seen
best by comparing the median concordance statistics (Table
\ref{tab:results}). Median concordance statistics in this simulation
scenario ranged from {[}0.79, 0.85{]}.\\
\strut \\
Finally, with strong class imbalance (event fraction = 0.02), the
effects of the imbalance corrections on discrimination were again,
highly dependent on the machine learning algorithm (Figure
\ref{fig:plot6}). For LR, imbalance corrections had no noticeable effect
on discrimination, with the exception of RUS, which decreased
discrimination (Figure \ref{fig:plot6}). For SVM and RF, all imbalance
corrections improved discrimination, meanwhile, for XG, RB, and EE, all
imbalance corrections worsened discrimination (Figure \ref{fig:plot6}).
Median concordance statistics in this simulation scenario ranged from
{[}0.71, 0.84{]}.\\
\strut \\
Overall, as imbalance between the classes was magnified, discrimination
worsened for most prediction models, on average. The effect of the
(pre-processing) imbalance corrections on model discrimination was
highly dependent on the machine learning algorithm. SVM benefited from
all imbalance corrections, RF also benefited, but to a lesser extent.
Meanwhile, XG, RB, EE and RUS-LR all suffered. Effects were most
pronounced when class imbalance was strong (event fraction = 0.02).
Notably, for any event fraction, discrimination for the machine learning
algorithms which innately corrected for imbalance (RB, EE), was worse
than the control-LR model.

\hypertarget{overall-performance}{%
\subsection{Overall Performance}\label{overall-performance}}

As Brier score depends on event fraction, we only compared Brier scores
for scenarios with the same event fraction. For reference, we note that
a trivial majority classifier (a model which produces the same predicted
risk, 0, for all individuals) would achieve Brier scores of 0.50, 0.19
and 0.02 in scenarios 4, 5, and 6, respectfully.\\
\strut \\
For balanced data (event fraction = 0.5), median Brier scores ranged
from {[}0.15, 0.19{]}. For models built with LR or RF, imbalance
corrections had no noticeable effect on median Brier score (Figure
\ref{fig:plot4}; Table \ref{tab:results}). For models built with SVM,
XG, RB or EE, imbalance corrections had minimal effects; the largest
difference in median Brier scores among prediction models built with the
same machine learning algorithm was \(0.1\) (Table \ref{tab:results}).\\
\strut \\
For moderately imbalanced data (event fraction = 0.2), median Brier
scores ranged from {[}0.12, 0.20{]}. Notably, the models which did not
attempt to correct for imbalance (control models for LR, SVM, RF and XG)
outperformed all other prediction models, with respect to Brier score
(Brier scores: 0.12). Again, with the exception of ROS-RF, which
preformed equally as well as the aforementioned models (Brier Score:
0.12). In this scenario, models built with RUS had particularly poor
performance relative to the others (Figure \ref{fig:plot5}); all RUS
models had median brier scores of 0.2, worse than a trivial majority
classifier, with the exception of RUS-SVM which preformed slightly
better (Brier Score: 0.16).\\
\strut \\
For data with strong imbalance (event fraction = 0.02), median Brier
scores ranged from {[}0.02, 0.22{]}. Again, the models which did not
attempt to correct for imbalance (control models for LR, SVM, RF and XG)
outperformed all other prediction models (Brier scores: 0.02). In this
scenario, ROS-RF and ROS-XG preformed just as well as the aforementioned
models (Brier scores: 0.02). Models built with RUS had substantially
worse Brier scores (ranging from {[}0.17, 0.22{]}) and with very large
MCMC errors, compared to all other models (Figure \ref{fig:plot6}).
Finally, while imbalance corrections worsened median Brier score for all
algorithms, the effect was most pronounced for LR (control: 0.02, RUS:
0.19, ROS: 0.16, SMOTE: 0.15, SENN: 0.16).\\
\strut \\
Overall, imbalance corrections worsened the overall performance for all
machine learning algorithms, on average. Two models were robust against
this effect: ROS-RF, ROS-XG. The effects of the imbalance corrections on
overall performance were most pronounced when class imbalance was
strong. Notably, the machine learning algorithms which innately
corrected for class imbalance (RB, EE) conferred no noticeable benefit
to overall performance compared to all other machine learning
algorithms.

\hypertarget{re-calibration}{%
\subsection{Re-calibration}\label{re-calibration}}

The effect of re-calibration on model performance was constant across
all simulation scenarios, and prediction models. Importantly, we
observed that even after re-calibration, there were very few cases
where, for a given machine learning algorithm, models trained with
imbalance corrected data preformed better than when trained with
uncorrected data, with respect to model calibration. These cases were
restricted to RUS-(SVM, RF) and ROS-RF models, in scenarios with
moderate (event fraction = 0.2) or strong class imbalance (event
fraction =0.02).\\
\strut \\
The re-calibration procedure adjusted calibration intercepts to be zero
for all simulation iterations, while calibration slopes were unaffected.
Consequently, after re-calibration, the flexible calibration plots
improved: there was less variability visible among the calibration
curves and less over-estimation of predicted risks. For the class
balanced scenarios, the separation seen among the calibration curves
disappeared. This can be viewed clearly using our Shiny App. With
respect to discrimination, concordance statistics were relatively
unaffected by re-calibration (Table \ref{tab:results}). Finally,
differences among the prediction models with respect to Brier score were
minimized after re-calibration such that all prediction models yielded
comparable Brier scores (Table \ref{tab:results}).\\
\strut \\
In summary, in scenarios where control-RF or control-SVM under-estimated
risk, RUS-(SVM, RF) and ROS-RF models slightly improved model
calibration, after re-calibration. This was most commonly observed in
simulation scenarios with 16 predictors (see Supplementary Materials
Section C).

\hypertarget{mimic-iii-data-case-study}{%
\section{MIMIC-III Data Case Study}\label{mimic-iii-data-case-study}}

\begin{figure}

{\centering \includegraphics[width=0.9\linewidth]{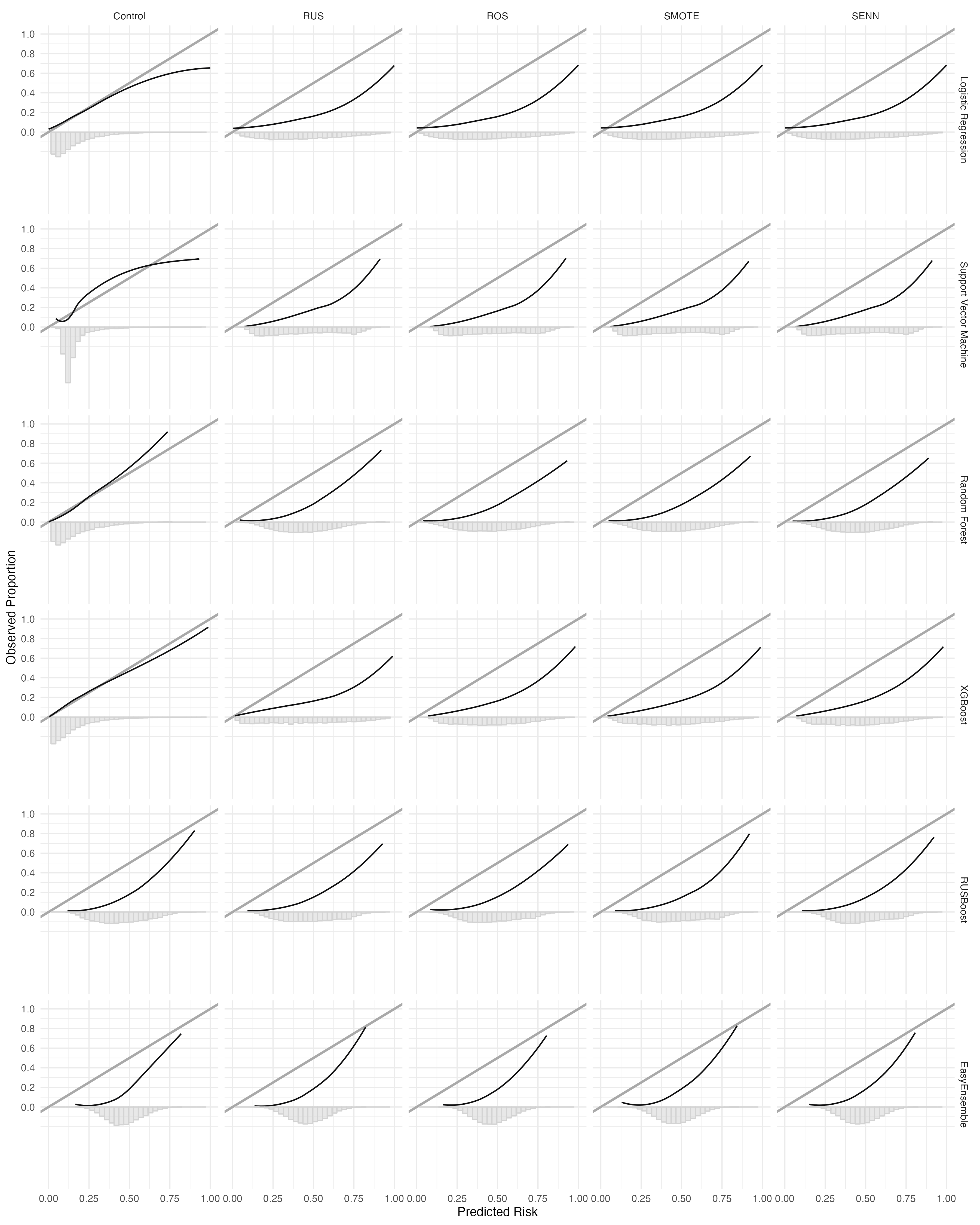} 

}

\caption{Flexible calibration curves for predictions models developed using the MIMIC-III database with histograms of the predicted risks.  The development data are characterized by 13 predictors, exactly the minimum required sample size and an event fraction of 0.17. Imbalance corrections: RUS (random undersampling), ROS (random oversampling), SMOTE (synthetic minority oversampling), SENN (synthetic minority oversampling with Wilson's Edited Nearest Neighbor rule). \label{fig:imp_ex_res}}\label{fig:unnamed-chunk-8}
\end{figure}

We conducted a case study to investigate how our simulation study
results compare with prediction models developed using empirical rather
than simulated data. Using the freely accessible MIMIC-III database
\cite{1-mimic_iii, 2-mimic_iii} we developed prediction models
predicting ICU mortality using commonly assessed predictors. We
developed and validated 30 (5 x 6) prediction models with the same
methods applied in our simulation study to assess the impact of 5
imbalance corrections (\(1\) control and \(4\) corrections) on
out-of-sample predictive performance for 6 different machine learning
algorithms.\\
\strut \\
The MIMIC-III database contains observations from 38,597 adult patients
admitted to the ICU at Beth Israel Deaconess Medical Center in Boston,
Massachusetts between 2008 and 2014 \cite{1-mimic_iii}. Many patients in
the database had more than one admission to the ICU, in our analysis
only data from the most recent admission was used. The outcome in our
analysis was death within 90 days of ICU admission. All predictors were
collected within the first 24 hrs of admission. For measurements taken
more than once within the first 24hrs of admission, the maximum value
was used, except for Glasgow coma score, where the minimum value was
used. We selected 13 predictors including, age, Glasgow coma score,
glucose, creatinine, hematocrit, hemoglobin, potassium, sodium, white
blood cell count, heart rate, mean blood pressure, respiratory rate and
temperature. We conducted a complete case analysis (as this study is
only meant as an illustration), and removed records with suspected
errors (two patients had a date of death recorded prior to their date of
admission and 1,890 patients had a recorded age of over 300 years). This
resulted in observations from 34,098 patients. We randomly split these
data into two sets that resembled development and validation settings.
All prediction models were developed using the same development data
(sample size 976 with 166 events, the minimum required sample size) and
the remainder of the data were used for validation (sample size 33,122
with 5567 events). The minimum required sample size calculation was
based on the observed event fraction (0.17), number of predictors (13)
and c-statistic (0.75), using the R package pmsampsize
\cite{pmsampsize}.\\
\strut \\
In Figure \ref{fig:imp_ex_res} the calibration plots for all 30
prediction models are presented, empirical performance metrics are
included in Appendix C. From Figure \ref{fig:imp_ex_res} it is clear
that in this case study, correcting for imbalance via a data
pre-processing technique (RUS, ROS, SMOTE, SENN), an algorithm which
innately corrects for class imbalance (RB, EE) or both, resulted in
prediction models which overestimated risk. With respect to
discrimination, all data pre-processing techniques used to correct for
class imbalance (RUS, ROS, SMOTE, SENN) worsened discrimination for each
machine learning algorithm considered except for support vector machine,
for which a slight benefit was conferred (Appendix C). The model with
the best discrimination was the control model for RF (Appendix C).
Finally, all prediction models developed with a class imbalance
correction had dramatically deteriorated overall performance as measured
by the Brier score (Appendix C).\\
\strut \\
Overall, these results bear striking similarity to that of simulation
scenario 5. In our simulation study, scenario 5 was characterized by
data with similar characteristics to the MIMIC-III data (event fraction
= 0.2, number of predictors = 8, sample size = minimum required sample
size). One notable difference was observed: in this example the random
forest model developed with oversampled data did exhibit miscalibration
(Figure \ref{fig:imp_ex_res}) whereas the random forest models developed
with oversampled data in simulation scenario 5 did not exhibit
miscalibration (Figure \ref{fig:plot5}).

\hypertarget{discussion}{%
\section{Discussion}\label{discussion}}

In this paper, we studied the impact of class imbalance corrections on
the out-of-sample predictive performance of clinical prediction models
developed with a variety of machine learning algorithms. We found that
when data exhibited class imbalance, implementing imbalance corrections
often led to deteriorated model calibration and (consequently)
deteriorated overall performance. For both moderate (event fraction =
0.2) and strong imbalance scenarios (event fraction = 0.02), we found
that correcting for class imbalance with a data pre-processing technique
(RUS, ROS, SMOTE, SENN) and/or an imbalance correcting algorithm (RB,
EE) resulted in prediction models that consistently over-estimated risk.
We noted one exception, ROS-RF models were often well calibrated in our
simulation study. Furthermore, the effect of imbalance corrections on
model discrimination was often small (or null) and highly dependent on
the machine learning algorithm. Subsequent re-calibration of predicted
risks improved model calibration and overall performance, yet, had no
effect on discrimination. Even after re-calibration, there were few
instances where models developed with an imbalance correction preformed
better than models developed with no correction for class imbalance. Our
findings were supported by the results of the case study. When
prediction models were developed using the MIMIC-III data, all
prediction models which corrected for class imbalance (including ROS-RF)
exhibited miscalibration characterized by an on average overestimation
of risk.\\
\strut \\
Our findings are consistent with those of van den Goorbergh
\emph{et al}. (2022) \cite{ruben}. In our simulation, pre-processing the
data to correct for class imbalance had no noticeable benefit for model
discrimination, and led to worse model calibration for all logistic
regression models. For prediction models developed with logistic
regression, we agree that imbalance corrections may do more harm than
good. This finding does, however, not hold for every machine learning
algorithm. In particular, for models developed with SVM or RF, the
findings were more nuanced. While imbalance corrections did improve
discrimination slightly, especially when data exhibited strong imbalance
(event fraction = 0.02), this often came at the cost of model
calibration. Using imbalanced corrected data to train SVM or RF models
resulted in deteriorated model calibration that was often not able to be
recovered even after re-calibration. After re-calibration, we noted
three models where imbalance corrections conferred a slight benefit to
both model discrimination and calibration in some simulation scenarios:
RUS-(SVM, RF) and ROS-RF. Finally, while previous studies found that
imbalance correcting algorithms improved model discrimination
\cite{kaur, rusboost, ee}, we found these algorithms (RB, EE), on
average conferred no benefit or worsened both model discrimination and
calibration in the 8 predictor scenarios. With 16 predictors, RB and EE
did show improved discrimination compared to the algorithms which did
not innately correct for imbalance, often at the cost of model
calibration.\\
\strut \\
Our study had two limitations which warrant discussion. Firstly,
hyperparameter tuning for SVM and XG was not as extensive as it was for
RF. More extensive tuning could perhaps improve model performance in
general, but would be unlikely to alter the effects seen as a
consequence of the imbalance corrections. Extensive tuning was
implemented for RF based on guidance from Probst \emph{et al}. (2019)
\cite{probst}, no such guidance was available for SVM and XG. Secondly,
our work focused on low dimensional settings (8 or 16 predictors) which
are typical for clinical prediction model development. Future research
may assess the impact of imbalance corrections on machine learning
models in higher dimensional settings. Additionally, the performance of
the random forest algorithm in combination with random oversampling was
an exception in our simulation study. Models developed with random
oversampling and random forest were often well-calibrated in our
simulation study, though not in our case study. Further research may
focus on the performance of random forest in combination with random
oversampling to better understand this finding.\\
\strut \\
Based on our findings we offer three considerations for researchers
developing prediction models in the presence of class imbalance. (1) We
found that correcting for class imbalance (with a data pre-processing
technique and/or an imbalance correcting algorithm) compromised model
calibration, resulting in prediction models which over-estimated risk,
in low-dimensional settings. Notably, the miscalibration conferred by
these corrections was often not restored by re-calibration and not
accompanied by improved discrimination. (2) After re-calibration, we
found that using random undersampling to pre-process training data for
support vector machine and random forest models sometimes conferred
slight benefit to predictive performance. In this case, we encourage
researchers to consider the ethical implications of discarding a
(potentially large) proportion of available data for what may be a small
gain in predictive performance. (3) Finally, we wish to highlight that
the machine learning algorithms which did not innately correct for class
imbalance (logistic regression, support vector machine, random forest
and XGBoost), were often well-calibrated when trained with imbalanced
data in scenarios with moderate imbalance (event fraction = 0.2). These
findings support the notion that it is not always necessary, and may
indeed be harmful, to correct for class imbalance.

\hypertarget{conclusion}{%
\section{Conclusion}\label{conclusion}}

Data exhibiting class imbalance are common in medical settings when the
modeled outcome is rare. Correcting for class imbalance is common, yet
little attention is paid to the effect of correcting for imbalance on
model calibration. If the goal of a clinical prediction model is to
produce reliable risk estimates (i.e., to achieve good calibration),
correcting for class imbalance with a data pre-processing technique
and/or an imbalance correcting algorithm may do more harm than good.

\newpage

\begin{table}[!h]

\caption{\label{tab:sim_sets} Summary of parameters for each of the 18 data-generating scenarios.}
\centering
\begin{tabular}[t]{ccccccccc}
\toprule
Scenario & No. Predictors & Sample Size & Event Fraction & $\mathrm{n}_{\mathrm{train}}$ & $\mathrm{n}_{\mathrm{validation}}$ & $\delta_\mu$ & $\delta_\Sigma$ & C\\
\midrule
1 & 8 & 0.5N & 0.50 & 193 & 1930 & 0.6043 & 0.3 & 0.85\\
2 & 8 & 0.5N & 0.20 & 124 & 1240 & 0.6043 & 0.3 & 0.85\\
3 & 8 & 0.5N & 0.02 & 899 & 8990 & 0.6043 & 0.3 & 0.85\\
4 & 8 & N & 0.50 & 385 & 3850 & 0.6043 & 0.3 & 0.85\\
5 & 8 & N & 0.20 & 247 & 2470 & 0.6043 & 0.3 & 0.85\\
6 & 8 & N & 0.02 & 1797 & 17970 & 0.6043 & 0.3 & 0.85\\
7 & 8 & 2N & 0.50 & 770 & 7700 & 0.6043 & 0.3 & 0.85\\
8 & 8 & 2N & 0.20 & 494 & 4940 & 0.6043 & 0.3 & 0.85\\
9 & 8 & 2N & 0.02 & 3594 & 35940 & 0.6043 & 0.3 & 0.85\\
10 & 16 & 0.5N & 0.50 & 193 & 1930 & 0.4854 & 0.3 & 0.85\\
11 & 16 & 0.5N & 0.20 & 247 & 2470 & 0.4854 & 0.3 & 0.85\\
12 & 16 & 0.5N & 0.02 & 1797 & 17970 & 0.4854 & 0.3 & 0.85\\
13 & 16 & N & 0.50 & 385 & 3850 & 0.4854 & 0.3 & 0.85\\
14 & 16 & N & 0.20 & 493 & 4930 & 0.4854 & 0.3 & 0.85\\
15 & 16 & N & 0.02 & 3593 & 35930 & 0.4854 & 0.3 & 0.85\\
16 & 16 & 2N & 0.50 & 770 & 7700 & 0.4854 & 0.3 & 0.85\\
17 & 16 & 2N & 0.20 & 986 & 9860 & 0.4854 & 0.3 & 0.85\\
18 & 16 & 2N & 0.02 & 7186 & 71860 & 0.4854 & 0.3 & 0.85\\
\bottomrule
\multicolumn{9}{l}{\rule{0pt}{1em}\textsuperscript{*} N: the minimum required sample size for a prediction model.}\\
\multicolumn{9}{l}{\rule{0pt}{1em}\textsuperscript{*} $\delta_{\mu}$: the difference in predictor means between the classes, for all predictors.}\\
\multicolumn{9}{l}{\rule{0pt}{1em}\textsuperscript{*} $\delta_{\Sigma}$: the difference in predictor variances between the classes, for all predictors.}\\
\multicolumn{9}{l}{\rule{0pt}{1em}\textsuperscript{*} $C$: the data-generating concordance statistic.}\\
\end{tabular}
\end{table}

\newpage

\begin{table}[!h]

\caption{\label{tab:corrections} Summary of imbalance corrections included in the simulation study.}
\centering
\begin{tabular}[t]{llll}
\toprule
Method & Abbreviation & Hyperparameters & R Package\\
\midrule
Random Undesampling & RUS &  & ROSE \cite{rose}\\
\addlinespace
Random Oversampling & ROS &  & ROSE \cite{rose}\\
\addlinespace
Synthetic Minority Over Sampling & SMOTE & $k=5$ & IRIC \cite{iric}\\
\addlinespace
SMOTE - Edited Nearest Neighbours & SENN & $k1=5$, $k2 = 3$ & IRIC \cite{iric}\\
\bottomrule
\multicolumn{4}{l}{\rule{0pt}{1em}\textsuperscript{*      $k$: the number of nearest neighbors in implementation of SMOTE.}}\\
\multicolumn{4}{l}{\rule{0pt}{1em}\textsuperscript{*    $k1$: the number of nearest neighbors in the SMOTE step of SENN.}}\\
\multicolumn{4}{l}{\rule{0pt}{1em}\textsuperscript{*     $k2$: the number of nearest neighbors in the ENN step of SENN.}}\\
\end{tabular}
\end{table}

\newpage

\begin{table}[!h]

\caption{\label{tab:algs}Summary of machine learning algorithms studied in the simulation study.}
\centering
\resizebox{\linewidth}{!}{
\begin{tabular}[t]{llll}
\toprule
Method & Abbreviation & Hyperparameter Tuning Grid & R Package\\
\midrule
Logistic Regression & LR &  & base R \cite{r}\\
\addlinespace
Support Vector Machine & SVM & default grid search & caret \cite{caret}\\
\addlinespace
Random Forest & RF & mtry [1: all predictors], min.node.size [1:10], splitrule [gini] & caret \cite{caret}\\
\addlinespace
XGBoost & XG & default grid search & caret \cite{caret}\\
\addlinespace
RUSBoost & RB &  & ebmc \cite{ebmc}\\
\addlinespace
EasyEnsemble & EE &  & IRIC \cite{iric}\\
\bottomrule
\multicolumn{4}{l}{\rule{0pt}{1em}\textsuperscript{* mtry[1: all predictors]: indicates that the number of candidate splitting variables may take on any value from 1 to the total number of predictors.}}\\
\multicolumn{4}{l}{\rule{0pt}{1em}\textsuperscript{* min.node.size[1: 10]: indicates that the minimum number of observations allowed in a leaf node may take on any integer value from 1 to 10.}}\\
\multicolumn{4}{l}{\rule{0pt}{1em}\textsuperscript{* splitrule[gini]: indicates that all random forest models select the split which minimizes Gini impurity.}}\\
\end{tabular}}
\end{table}

\begin{landscape}
\begin{table}
\caption{\label{tab:results} Median performance measures and their Monte Carlo errors across 2000 simulation iterations for simulation scenarios 4-6.}
\centering
\resizebox{\linewidth}{!}{
\begin{tabular}[t]{lrrrrrrrrrrrrrrrrrrrrrrrrrrrrrr}
\toprule
\multicolumn{1}{c}{ } & \multicolumn{6}{c}{Control} & \multicolumn{6}{c}{RUS} & \multicolumn{6}{c}{ROS} & \multicolumn{6}{c}{SMOTE} & \multicolumn{6}{c}{SENN} \\
\cmidrule(l{3pt}r{3pt}){2-7} \cmidrule(l{3pt}r{3pt}){8-13} \cmidrule(l{3pt}r{3pt}){14-19} \cmidrule(l{3pt}r{3pt}){20-25} \cmidrule(l{3pt}r{3pt}){26-31}
  & LR & SVM & RF & XG & RB & EE & LR & SVM & RF & XG & RB & EE & LR & SVM & RF & XG & RB & EE & LR & SVM & RF & XG & RB & EE & LR & SVM & RF & XG & RB & EE\\
\midrule
\addlinespace[0.3em]
\multicolumn{31}{l}{\textbf{Scenario 4}}\\
\hspace{2em}\hspace{1em}Concordance Statistic & 0.84 & 0.86 & 0.84 & 0.84 & 0.84 & 0.83 & 0.84 & 0.86 & 0.84 & 0.84 & 0.83 & 0.82 & 0.84 & 0.85 & 0.84 & 0.82 & 0.83 & 0.82 & 0.84 & 0.86 & 0.84 & 0.84 & 0.83 & 0.82 & 0.84 & 0.85 & 0.84 & 0.84 & 0.83 & 0.83\\
\hspace{2em}\hspace{1em}MCMC Error & 0.01 & 0.01 & 0.01 & 0.01 & 0.01 & 0.01 & 0.01 & 0.01 & 0.01 & 0.01 & 0.01 & 0.01 & 0.01 & 0.01 & 0.01 & 0.01 & 0.01 & 0.01 & 0.01 & 0.01 & 0.01 & 0.01 & 0.01 & 0.01 & 0.01 & 0.01 & 0.01 & 0.01 & 0.01 & \vphantom{1} 0.01\\
\hspace{2em}\hspace{1em}Brier Score & 0.16 & 0.15 & 0.16 & 0.16 & 0.17 & 0.19 & 0.16 & 0.15 & 0.16 & 0.16 & 0.17 & 0.19 & 0.16 & 0.16 & 0.17 & 0.19 & 0.17 & 0.19 & 0.16 & 0.15 & 0.16 & 0.16 & 0.17 & 0.19 & 0.17 & 0.17 & 0.17 & 0.18 & 0.17 & 0.17\\
\hspace{2em}\hspace{1em}MCMC Error & <0.01 & <0.01 & <0.01 & 0.01 & <0.01 & <0.01 & <0.01 & <0.01 & <0.01 & 0.01 & 0.01 & <0.01 & <0.01 & 0.01 & 0.01 & 0.01 & 0.01 & 0.01 & <0.01 & <0.01 & <0.01 & 0.01 & 0.01 & <0.01 & 0.01 & 0.01 & 0.01 & 0.01 & 0.01 & 0.01\\
\hspace{2em}\hspace{1em}Calibration Intercept & <0.01 & <0.01 & <0.01 & <0.01 & <0.01 & <0.01 & <0.01 & <0.01 & <0.01 & <0.01 & <0.01 & <0.01 & <0.01 & <0.01 & <0.01 & <0.01 & <0.01 & <0.01 & <0.01 & <0.01 & <0.01 & <0.01 & <0.01 & <0.01 & <0.01 & <0.01 & <0.01 & <0.01 & <0.01 & \vphantom{5} <0.01\\
\hspace{2em}\hspace{1em}MCMC Error & 0.14 & 0.14 & 0.12 & 0.15 & 0.07 & 0.05 & 0.11 & 0.12 & 0.09 & 0.11 & 0.07 & 0.05 & 0.13 & 0.20 & 0.34 & 0.51 & 0.18 & 0.17 & 0.10 & 0.11 & 0.09 & 0.11 & 0.08 & 0.05 & 0.23 & 0.25 & 0.15 & 0.26 & 0.14 & 0.07\\
\hspace{2em}\hspace{1em}Calibration Slope & 0.92 & 0.96 & 1.25 & 0.91 & 1.46 & >10.0 & 0.92 & 0.96 & 1.25 & 0.90 & 1.46 & >10.0 & 0.88 & 0.87 & 1.21 & 0.56 & 1.26 & >10.0 & 0.90 & 0.94 & 1.21 & 0.89 & 1.43 & >10.0 & 0.57 & 0.58 & 0.78 & 0.52 & 0.83 & 1.62\\
\hspace{2em}\hspace{1em}MCMC Error & 0.10 & 0.11 & 0.18 & 0.12 & 0.17 & 0.17 & 0.11 & 0.11 & 0.18 & 0.12 & 0.17 & 0.17 & 0.12 & 0.11 & 0.19 & 0.10 & 0.15 & 0.16 & 0.10 & 0.11 & 0.18 & 0.12 & 0.16 & 0.16 & 0.19 & 0.20 & 0.27 & 0.21 & 0.33 & 0.34\\
\addlinespace \addlinespace[0.3em]
\multicolumn{31}{l}{\textbf{Scenario 5}}\\
\hspace{2em}\hspace{1em}Concordance Statistic & 0.84 & 0.82 & 0.82 & 0.82 & 0.83 & 0.82 & 0.83 & 0.85 & 0.83 & 0.80 & 0.81 & 0.80 & 0.83 & 0.83 & 0.83 & 0.79 & 0.80 & 0.80 & 0.84 & 0.83 & 0.82 & 0.81 & 0.80 & 0.81 & 0.83 & 0.83 & 0.82 & 0.82 & 0.81 & 0.82\\
\hspace{2em}\hspace{1em}MCMC Error & 0.01 & 0.03 & 0.02 & 0.02 & 0.02 & 0.02 & 0.02 & 0.01 & 0.02 & 0.02 & 0.02 & 0.02 & 0.01 & 0.02 & 0.02 & 0.02 & 0.02 & 0.02 & 0.01 & 0.02 & 0.02 & 0.02 & 0.02 & 0.02 & 0.01 & 0.02 & 0.02 & 0.02 & 0.02 & 0.02\\
\hspace{2em}\hspace{1em}Brier Score & 0.12 & 0.12 & 0.12 & 0.12 & 0.17 & 0.20 & 0.18 & 0.16 & 0.18 & 0.19 & 0.20 & 0.20 & 0.17 & 0.15 & 0.12 & 0.15 & 0.15 & 0.15 & 0.16 & 0.14 & 0.13 & 0.15 & 0.14 & 0.15 & 0.19 & 0.16 & 0.15 & 0.17 & 0.16 & 0.15\\
\hspace{2em}\hspace{1em}MCMC Error & 0.01 & 0.01 & 0.01 & 0.01 & 0.01 & 0.01 & 0.02 & 0.02 & 0.02 & 0.02 & 0.02 & 0.01 & 0.01 & 0.01 & 0.01 & 0.01 & 0.01 & 0.01 & 0.01 & 0.01 & 0.01 & 0.01 & 0.01 & 0.01 & 0.02 & 0.01 & 0.01 & 0.01 & 0.01 & 0.01\\
\hspace{2em}\hspace{1em}Calibration Intercept & <0.01 & <0.01 & <0.01 & <0.01 & <0.01 & <0.01 & <0.01 & <0.01 & <0.01 & <0.01 & <0.01 & <0.01 & <0.01 & <0.01 & <0.01 & <0.01 & <0.01 & <0.01 & <0.01 & <0.01 & <0.01 & <0.01 & <0.01 & <0.01 & <0.01 & <0.01 & <0.01 & <0.01 & <0.01 & \vphantom{4} <0.01\\
\hspace{2em}\hspace{1em}MCMC Error & 0.22 & 0.23 & 0.18 & 0.23 & 0.10 & 0.07 & 0.40 & 0.23 & 0.18 & 0.36 & 0.14 & 0.08 & 0.16 & 0.24 & 0.17 & 0.43 & 0.17 & 0.10 & 0.18 & 0.30 & 0.18 & 0.33 & 0.20 & 0.10 & 0.30 & 0.40 & 0.24 & 0.46 & 0.25 & 0.10\\
\hspace{2em}\hspace{1em}Calibration Slope & 0.85 & 0.97 & 1.27 & 0.79 & 1.52 & >10.0 & 0.67 & 0.96 & 1.26 & 0.62 & 1.30 & >10.0 & 0.76 & 0.61 & 1.10 & 0.33 & 0.85 & 1.78 & 0.71 & 0.55 & 0.84 & 0.41 & 0.78 & 1.71 & 0.54 & 0.43 & 0.66 & 0.35 & 0.58 & 1.48\\
\hspace{2em}\hspace{1em}MCMC Error & 0.14 & 0.55 & 0.20 & 0.13 & 0.24 & 0.27 & 0.17 & 0.26 & 0.31 & 0.14 & 0.24 & 0.27 & 0.14 & 0.11 & 0.20 & 0.05 & 0.13 & 0.18 & 0.13 & 0.10 & 0.15 & 0.07 & 0.13 & 0.17 & 0.12 & 0.08 & 0.15 & 0.05 & 0.10 & 0.16\\
\addlinespace \addlinespace[0.3em]
\multicolumn{31}{l}{\textbf{Scenario 6}}\\
\hspace{2em}\hspace{1em}Concordance Statistic & 0.84 & 0.71 & 0.78 & 0.82 & 0.83 & 0.81 & 0.82 & 0.84 & 0.82 & 0.79 & 0.81 & 0.79 & 0.84 & 0.80 & 0.81 & 0.76 & 0.76 & 0.76 & 0.84 & 0.80 & 0.80 & 0.78 & 0.76 & 0.78 & 0.84 & 0.80 & 0.80 & 0.79 & 0.77 & 0.79\\
\hspace{2em}\hspace{1em}MCMC Error & 0.01 & 0.04 & 0.02 & 0.02 & 0.02 & 0.02 & 0.03 & 0.02 & 0.02 & 0.03 & 0.03 & 0.03 & 0.01 & 0.02 & 0.02 & 0.02 & 0.03 & 0.02 & 0.01 & 0.02 & 0.02 & 0.02 & 0.02 & 0.02 & 0.01 & 0.02 & 0.02 & 0.02 & 0.02 & 0.02\\
\hspace{2em}\hspace{1em}Brier Score & 0.02 & 0.02 & 0.02 & 0.02 & 0.16 & 0.20 & 0.19 & 0.17 & 0.18 & 0.21 & 0.22 & 0.20 & 0.16 & 0.05 & 0.02 & 0.02 & 0.04 & 0.04 & 0.15 & 0.05 & 0.03 & 0.03 & 0.03 & 0.05 & 0.16 & 0.05 & 0.03 & 0.04 & 0.04 & 0.05\\
\hspace{2em}\hspace{1em}MCMC Error & <0.01 & <0.01 & <0.01 & <0.01 & 0.02 & 0.01 & 0.04 & 0.03 & 0.03 & 0.04 & 0.03 & 0.02 & 0.01 & 0.01 & <0.01 & <0.01 & <0.01 & <0.01 & 0.02 & 0.01 & <0.01 & <0.01 & <0.01 & 0.01 & 0.02 & 0.01 & <0.01 & <0.01 & <0.01 & 0.01\\
\hspace{2em}\hspace{1em}Calibration Intercept & <0.01 & <0.01 & <0.01 & <0.01 & <0.01 & <0.01 & <0.01 & <0.01 & <0.01 & <0.01 & <0.01 & <0.01 & <0.01 & <0.01 & <0.01 & <0.01 & <0.01 & <0.01 & <0.01 & <0.01 & <0.01 & <0.01 & <0.01 & <0.01 & <0.01 & <0.01 & <0.01 & <0.01 & <0.01 & \vphantom{3} <0.01\\
\hspace{2em}\hspace{1em}MCMC Error & <0.01 & <0.01 & <0.01 & <0.01 & <0.01 & <0.01 & <0.01 & <0.01 & <0.01 & <0.01 & <0.01 & <0.01 & <0.01 & <0.01 & >10.0 & <0.01 & <0.01 & <0.01 & <0.01 & <0.01 & <0.01 & <0.01 & <0.01 & <0.01 & <0.01 & <0.01 & <0.01 & <0.01 & <0.01 & <0.01\\
\hspace{2em}\hspace{1em}Calibration Slope & 0.90 & 1.41 & 0.63 & 0.84 & 1.49 & >10.0 & 0.54 & 0.96 & 1.18 & 0.52 & 1.26 & 1.86 & 0.72 & 0.14 & 0.53 & 0.25 & 0.35 & 1.67 & 0.62 & 0.16 & 0.48 & 0.22 & 0.28 & 1.50 & 0.57 & 0.15 & 0.45 & 0.20 & 0.25 & 1.38\\
\hspace{2em}\hspace{1em}MCMC Error & 0.12 & >10.0 & 0.20 & 0.13 & 0.28 & 0.30 & 0.18 & 0.66 & 0.32 & 0.12 & 0.26 & 0.28 & 0.13 & 0.04 & 0.18 & 0.02 & 0.06 & 0.17 & 0.11 & 0.03 & 0.12 & 0.02 & 0.05 & 0.13 & 0.11 & 0.03 & 0.11 & 0.02 & 0.04 & 0.12\\
\addlinespace \addlinespace[0.3em]
\multicolumn{31}{l}{\textbf{Scenario 4 Recalibrated}}\\
\hspace{2em}\hspace{1em}Concordance Statistic & 0.84 & 0.84 & 0.83 & 0.82 & 0.84 & 0.85 & 0.84 & 0.82 & 0.83 & 0.82 & 0.84 & 0.86 & 0.86 & 0.84 & 0.84 & 0.83 & 0.82 & 0.84 & 0.85 & 0.84 & 0.84 & 0.83 & 0.84 & 0.83 & 0.84 & 0.84 & 0.83 & 0.84 & 0.86 & 0.84\\
\hspace{2em}\hspace{1em}MCMC Error & 0.01 & 0.01 & 0.01 & 0.01 & 0.01 & 0.01 & 0.01 & 0.01 & 0.01 & 0.01 & 0.01 & 0.01 & 0.01 & 0.01 & 0.01 & 0.01 & 0.01 & 0.01 & 0.01 & 0.01 & 0.01 & 0.01 & 0.01 & 0.01 & 0.01 & 0.01 & 0.01 & 0.01 & 0.01 & 0.01\\
\hspace{2em}\hspace{1em}Brier Score & 0.16 & 0.16 & 0.17 & 0.19 & 0.16 & 0.16 & 0.17 & 0.18 & 0.17 & 0.19 & 0.16 & 0.15 & 0.15 & 0.16 & 0.16 & 0.17 & 0.19 & 0.17 & 0.16 & 0.17 & 0.17 & 0.17 & 0.16 & 0.17 & 0.16 & 0.17 & 0.19 & 0.16 & 0.15 & 0.16\\
\hspace{2em}\hspace{1em}MCMC Error & <0.01 & 0.01 & 0.01 & <0.01 & <0.01 & 0.01 & <0.01 & 0.01 & 0.01 & 0.01 & <0.01 & <0.01 & <0.01 & <0.01 & 0.01 & 0.01 & <0.01 & 0.01 & 0.01 & 0.01 & 0.01 & 0.01 & <0.01 & 0.01 & <0.01 & <0.01 & <0.01 & <0.01 & <0.01 & <0.01\\
\hspace{2em}\hspace{1em}Calibration Intercept & <0.01 & <0.01 & <0.01 & <0.01 & <0.01 & <0.01 & <0.01 & <0.01 & <0.01 & <0.01 & <0.01 & <0.01 & <0.01 & <0.01 & <0.01 & <0.01 & <0.01 & <0.01 & <0.01 & <0.01 & <0.01 & <0.01 & <0.01 & <0.01 & <0.01 & <0.01 & <0.01 & <0.01 & <0.01 & \vphantom{2} <0.01\\
\hspace{2em}\hspace{1em}MCMC Error & <0.01 & <0.01 & <0.01 & <0.01 & <0.01 & <0.01 & <0.01 & <0.01 & <0.01 & <0.01 & <0.01 & <0.01 & <0.01 & <0.01 & <0.01 & <0.01 & <0.01 & <0.01 & <0.01 & <0.01 & <0.01 & <0.01 & <0.01 & <0.01 & <0.01 & <0.01 & <0.01 & <0.01 & <0.01 & \vphantom{2} <0.01\\
\hspace{2em}\hspace{1em}Calibration Slope & 0.92 & 0.90 & 1.46 & >10.0 & 0.88 & 0.87 & 1.21 & 0.56 & 1.26 & >10.0 & 0.90 & 0.96 & 0.94 & 1.21 & 0.89 & 1.43 & >10.0 & 0.57 & 0.58 & 0.78 & 0.52 & 0.83 & 1.25 & 1.62 & 0.91 & 1.46 & >10.0 & 0.92 & 0.96 & 1.25\\
\hspace{2em}\hspace{1em}MCMC Error & 0.10 & 0.12 & 0.17 & 0.17 & 0.12 & 0.11 & 0.19 & 0.10 & 0.15 & 0.16 & 0.10 & 0.11 & 0.11 & 0.18 & 0.12 & 0.16 & 0.16 & 0.19 & 0.20 & 0.27 & 0.21 & 0.33 & 0.18 & 0.34 & 0.12 & 0.17 & 0.17 & 0.11 & 0.11 & 0.18\\
\addlinespace \addlinespace[0.3em]
\multicolumn{31}{l}{\textbf{Scenario 5 Recalibrated}}\\
\hspace{2em}\hspace{1em}Concordance Statistic & 0.84 & 0.80 & 0.81 & 0.80 & 0.83 & 0.83 & 0.83 & 0.79 & 0.80 & 0.80 & 0.84 & 0.82 & 0.83 & 0.82 & 0.81 & 0.80 & 0.81 & 0.83 & 0.83 & 0.82 & 0.82 & 0.81 & 0.82 & 0.82 & 0.82 & 0.83 & 0.82 & 0.83 & 0.85 & 0.83\\
\hspace{2em}\hspace{1em}MCMC Error & 0.01 & 0.02 & 0.02 & 0.02 & 0.01 & 0.02 & 0.02 & 0.02 & 0.02 & 0.02 & 0.01 & 0.03 & 0.02 & 0.02 & 0.02 & 0.02 & 0.02 & 0.01 & 0.02 & 0.02 & 0.02 & 0.02 & 0.02 & 0.02 & 0.02 & 0.02 & 0.02 & 0.02 & 0.01 & 0.02\\
\hspace{2em}\hspace{1em}Brier Score & 0.12 & 0.13 & 0.13 & 0.14 & 0.12 & 0.13 & 0.12 & 0.15 & 0.13 & 0.13 & 0.12 & 0.12 & 0.13 & 0.12 & 0.14 & 0.13 & 0.13 & 0.13 & 0.14 & 0.13 & 0.15 & 0.13 & 0.12 & 0.13 & 0.12 & 0.12 & 0.13 & 0.13 & 0.12 & 0.12\\
\hspace{2em}\hspace{1em}MCMC Error & 0.01 & 0.01 & 0.01 & 0.01 & 0.01 & 0.01 & 0.01 & 0.01 & 0.01 & <0.01 & 0.01 & 0.01 & 0.01 & 0.01 & 0.01 & 0.01 & <0.01 & 0.01 & 0.01 & 0.01 & 0.01 & 0.01 & <0.01 & 0.01 & 0.01 & 0.01 & <0.01 & 0.01 & 0.01 & 0.01\\
\hspace{2em}\hspace{1em}Calibration Intercept & <0.01 & <0.01 & <0.01 & <0.01 & <0.01 & <0.01 & <0.01 & <0.01 & <0.01 & <0.01 & <0.01 & <0.01 & <0.01 & <0.01 & <0.01 & <0.01 & <0.01 & <0.01 & <0.01 & <0.01 & <0.01 & <0.01 & <0.01 & <0.01 & <0.01 & <0.01 & <0.01 & <0.01 & <0.01 & \vphantom{1} <0.01\\
\hspace{2em}\hspace{1em}MCMC Error & <0.01 & <0.01 & <0.01 & <0.01 & <0.01 & <0.01 & <0.01 & <0.01 & <0.01 & <0.01 & <0.01 & <0.01 & <0.01 & <0.01 & <0.01 & <0.01 & <0.01 & <0.01 & <0.01 & <0.01 & <0.01 & <0.01 & <0.01 & <0.01 & <0.01 & <0.01 & <0.01 & <0.01 & <0.01 & \vphantom{1} <0.01\\
\hspace{2em}\hspace{1em}Calibration Slope & 0.85 & 0.62 & 1.30 & >10.0 & 0.76 & 0.61 & 1.10 & 0.33 & 0.85 & 1.78 & 0.71 & 0.97 & 0.55 & 0.84 & 0.41 & 0.78 & 1.71 & 0.54 & 0.43 & 0.66 & 0.35 & 0.58 & 1.27 & 1.48 & 0.79 & 1.52 & >10.0 & 0.67 & 0.96 & 1.26\\
\hspace{2em}\hspace{1em}MCMC Error & 0.14 & 0.14 & 0.24 & 0.27 & 0.14 & 0.11 & 0.20 & 0.05 & 0.13 & 0.18 & 0.13 & 0.55 & 0.10 & 0.15 & 0.07 & 0.13 & 0.17 & 0.12 & 0.08 & 0.15 & 0.05 & 0.10 & 0.20 & 0.16 & 0.13 & 0.24 & 0.27 & 0.17 & 0.26 & 0.31\\
\addlinespace \addlinespace[0.3em]
\multicolumn{31}{l}{\textbf{Scenario 6 Recalibrated}}\\
\hspace{2em}\hspace{1em}Concordance Statistic & 0.84 & 0.79 & 0.81 & 0.79 & 0.84 & 0.79 & 0.80 & 0.76 & 0.76 & 0.76 & 0.84 & 0.71 & 0.80 & 0.80 & 0.78 & 0.76 & 0.78 & 0.84 & 0.80 & 0.80 & 0.79 & 0.77 & 0.78 & 0.79 & 0.82 & 0.83 & 0.81 & 0.82 & 0.84 & 0.82\\
\hspace{2em}\hspace{1em}MCMC Error & 0.01 & 0.03 & 0.03 & 0.03 & 0.01 & 0.03 & 0.10 & 0.02 & 0.03 & 0.02 & 0.01 & 0.04 & 0.02 & 0.04 & 0.02 & 0.02 & 0.02 & 0.01 & 0.02 & 0.05 & 0.02 & 0.02 & 0.03 & 0.02 & 0.02 & 0.02 & 0.02 & 0.03 & 0.02 & 0.02\\
\hspace{2em}\hspace{1em}Brier Score & 0.02 & 0.02 & 0.02 & 0.02 & 0.02 & 0.03 & 0.02 & 0.03 & 0.02 & 0.02 & 0.02 & 0.02 & 0.03 & 0.02 & 0.03 & 0.02 & 0.02 & 0.02 & 0.03 & 0.02 & 0.03 & 0.02 & 0.02 & 0.02 & 0.02 & 0.02 & 0.02 & 0.02 & 0.02 & 0.02\\
\hspace{2em}\hspace{1em}MCMC Error & <0.01 & <0.01 & <0.01 & <0.01 & <0.01 & <0.01 & <0.01 & <0.01 & <0.01 & <0.01 & <0.01 & <0.01 & <0.01 & <0.01 & <0.01 & <0.01 & <0.01 & <0.01 & <0.01 & <0.01 & <0.01 & <0.01 & <0.01 & <0.01 & <0.01 & <0.01 & <0.01 & <0.01 & <0.01 & <0.01\\
\hspace{2em}\hspace{1em}Calibration Intercept & <0.01 & <0.01 & <0.01 & <0.01 & <0.01 & <0.01 & <0.01 & <0.01 & <0.01 & <0.01 & <0.01 & <0.01 & <0.01 & <0.01 & <0.01 & <0.01 & <0.01 & <0.01 & <0.01 & <0.01 & <0.01 & <0.01 & <0.01 & <0.01 & <0.01 & <0.01 & <0.01 & <0.01 & <0.01 & <0.01\\
\hspace{2em}\hspace{1em}MCMC Error & <0.01 & <0.01 & <0.01 & <0.01 & <0.01 & <0.01 & >10.0 & <0.01 & <0.01 & <0.01 & <0.01 & <0.01 & <0.01 & <0.01 & <0.01 & <0.01 & <0.01 & <0.01 & <0.01 & <0.01 & <0.01 & <0.01 & <0.01 & <0.01 & <0.01 & <0.01 & <0.01 & <0.01 & <0.01 & <0.01\\
\hspace{2em}\hspace{1em}Calibration Slope & 0.90 & 0.52 & 1.26 & 1.86 & 0.72 & 0.14 & 0.56 & 0.25 & 0.35 & 1.67 & 0.62 & 1.41 & 0.16 & 0.48 & 0.22 & 0.28 & 1.50 & 0.57 & 0.15 & 0.45 & 0.20 & 0.25 & 0.63 & 1.38 & 0.84 & 1.49 & >10.0 & 0.54 & 0.96 & 1.18\\
\hspace{2em}\hspace{1em}MCMC Error & 0.12 & 0.12 & 0.26 & 0.28 & 0.13 & 0.05 & 0.16 & 0.02 & 0.06 & 0.17 & 0.11 & >10.0 & 0.03 & 0.12 & 0.02 & 0.05 & 0.13 & 0.11 & 0.03 & 0.11 & 0.02 & 0.04 & 0.20 & 0.12 & 0.13 & 0.28 & 0.30 & 0.18 & 0.66 & 0.32\\
\bottomrule
\multicolumn{31}{l}{\rule{0pt}{1em}\textsuperscript{1} Imbalance Corrections: RUS (random undersampling), ROS (random oversampling), SMOTE (synthetic minority oversampling), SENN (synthetic minority oversampling with Wilson's Edited Nearest Neighbor rule)}\\
\multicolumn{31}{l}{\rule{0pt}{1em}\textsuperscript{2} Machine Learning Algorithms: LR (logistic regression), SVM (support vector machine), RF (random forest), XG (XGBoost), RB (RUSBoost), EE (EasyEnsemble)}\\
\multicolumn{31}{l}{\rule{0pt}{1em}\textsuperscript{3} Scenario 4: 8 predictors, minimum required sample size (N), class balanced data (event fraction = 0.5)}\\
\multicolumn{31}{l}{\rule{0pt}{1em}\textsuperscript{4} Scenario 5: 8 predictors, minimum required sample size (N), moderately imbalanced data (event fraction = 0.2)}\\
\multicolumn{31}{l}{\rule{0pt}{1em}\textsuperscript{5} Scenario 6: 8 predictors, minimum required sample size (N), strongly imbalanced data (event fraction = 0.02)}\\
\end{tabular}}
\end{table}
\end{landscape}


\clearpage

\bibliographystyle{unsrt}
\bibliography{manuscript.bib}

\begin{thebibliography}{10}

\bibitem{ewout_intro}
E.W. Steyerberg.
\newblock {\em Applications of prediction models}, pages 11--31.
\newblock Springer New York, New York, NY, 2009.

\bibitem{annals}
Lingxiao Chen.
\newblock Overview of clinical prediction models.
\newblock {\em Annals of Translational Medicine}, 8(4), 2019.

\bibitem{achilles}
Ben Van~Calster, David~J. McLernon, Maarten van Smeden, Laure Wynants, Ewout~W. Steyerberg, Patrick Bossuyt, Gary~S. Collins, Petra Macaskill, David~J. McLernon, Karel G.~M. Moons, Ewout~W. Steyerberg, Andrew~J. Vickers, On~behalf of Topic Group `Evaluating~diagnostic tests, and prediction models'of~the STRATOS~initiative.
\newblock Calibration: the achilles heel of predictive analytics.
\newblock {\em BMC Medicine}, 17(1):230, 2019.

\bibitem{van_smeden_clinical_2021}
Maarten van Smeden, Johannes~B. Reitsma, Richard~D. Riley, Gary~S. Collins, and Karel~Gm Moons.
\newblock Clinical prediction models: diagnosis versus prognosis.
\newblock {\em Journal of Clinical Epidemiology}, 132:142--145, April 2021.

\bibitem{moons_prognosis_2009}
Karel G.~M. Moons, Douglas~G. Altman, Yvonne Vergouwe, and Patrick Royston.
\newblock Prognosis and prognostic research: application and impact of prognostic models in clinical practice.
\newblock {\em BMJ}, 338:b606, June 2009.
\newblock Publisher: British Medical Journal Publishing Group Section: Research Methods \&amp; Reporting.

\bibitem{cip}
Fadel~M. Megahed, Ying-Ju Chen, Aly Megahed, Yuya Ong, Naomi Altman, and Martin Krzywinski.
\newblock The class imbalance problem.
\newblock {\em Nature Methods}, 18(11):1270--1272, 2021.

\bibitem{summary_m}
Satyam Maheshwari, R.C. Jain, and R.S. Jadon.
\newblock An insight into rare class problem: Analysis and potential solutions.
\newblock {\em Journal of Computer Science}, 14(6):777--792, May 2018.

\bibitem{lp}
Victoria L{\'o}pez, Alberto Fern{\'a}ndez, Jose~G. Moreno-Torres, and Francisco Herrera.
\newblock Analysis of preprocessing vs. cost-sensitive learning for imbalanced classification. open problems on intrinsic data characteristics.
\newblock {\em Expert Systems with Applications}, 39(7):6585--6608, 2012.

\bibitem{summary_h}
Guo Haixiang, Li~Yijing, Jennifer Shang, Gu~Mingyun, Huang Yuanyue, and Gong Bing.
\newblock Learning from class-imbalanced data: Review of methods and applications.
\newblock {\em Expert Systems with Applications}, 73:220--239, 2017.

\bibitem{12days}
Richard~D Riley, Tim~J Cole, Jon Deeks, Jamie~J Kirkham, Julie Morris, Rafael Perera, Angie Wade, and Gary~S Collins.
\newblock On the 12th day of christmas, a statistician sent to me . . .
\newblock {\em BMJ}, 379, 2022.

\bibitem{bens_paper}
Ben~Van Calster and Andrew~J. Vickers.
\newblock Calibration of risk prediction models: Impact on decision-analytic performance.
\newblock {\em Medical Decision Making}, 35(2):162--169, 2015.

\bibitem{biesheuvel_advantages_2008}
Cornelis~J. Biesheuvel, Yvonne Vergouwe, Ruud Oudega, Arno~W. Hoes, Diederick~E. Grobbee, and Karel G.~M. Moons.
\newblock Advantages of the nested case-control design in diagnostic research.
\newblock {\em BMC medical research methodology}, 8:48, July 2008.

\bibitem{ruben}
Ruben van~den Goorbergh, Maarten van Smeden, Dirk Timmerman, and Ben Van~Calster.
\newblock {The harm of class imbalance corrections for risk prediction models: illustration and simulation using logistic regression}.
\newblock {\em Journal of the American Medical Informatics Association}, 29(9):1525--1534, 06 2022.

\bibitem{constanza}
Constanza Navarro, Johanna Damen, Maarten van Smeden, Toshihiko Takada, Steven Nijman, Paula Dhiman, Jie Ma, Gary Collins, Ram Bajpai, Richard Riley, Karel Moons, and Lotty Hooft.
\newblock Systematic review identifies the design and methodological conduct of studies on machine learning-based prediction models.
\newblock {\em Journal of Clinical Epidemiology}, 11 2022.

\bibitem{pmsampsize}
{Joie Ensor and Emma C. Martin and Richard D. Riley}.
\newblock {\em pmsampsize: Calculates the Minimum Sample Size Required for Developing a Multivariable Prediction Model}, 2022.
\newblock R package version 1.1.2.

\bibitem{mvauc}
Olga~V. Demler, Michael~J. Pencina, and Ralph~B. D'Agostino~Sr.
\newblock Equivalence of improvement in area under roc curve and linear discriminant analysis coefficient under assumption of normality.
\newblock {\em Statistics in Medicine}, 30(12):1410--1418, 2011.

\bibitem{senn}
Gustavo E. A. P.~A. Batista, Ronaldo~C. Prati, and Maria~Carolina Monard.
\newblock A study of the behavior of several methods for balancing machine learning training data.
\newblock {\em SIGKDD Explor. Newsl.}, 6(1):20–29, jun 2004.

\bibitem{heart_failure_senn}
Qianchuan Zhao, Mirza Muntasir~Nishat, Fahim Faisal, Ishrak Jahan~Ratul, Abdullah Al-Monsur, Abrar~Mohammad Ar-Rafi, Sarker~Mohammad Nasrullah, Md~Taslim Reza, and Md~Rezaul~Hoque Khan.
\newblock A comprehensive investigation of the performances of different machine learning classifiers with smote-enn oversampling technique and hyperparameter optimization for imbalanced heart failure dataset.
\newblock {\em Scientific Programming}, 2022:3649406, 2022.

\bibitem{rutjes_casecontrol_2005}
Anne~WS Rutjes, Johannes~B Reitsma, Jan~P Vandenbroucke, Afina~S Glas, and Patrick~MM Bossuyt.
\newblock Case–{Control} and {Two}-{Gate} {Designs} in {Diagnostic} {Accuracy} {Studies}.
\newblock {\em Clinical Chemistry}, 51:1335--1341, August 2005.
\newblock \_eprint: https://academic.oup.com/clinchem/article-pdf/51/8/1335/32682656/clinchem1335.pdf.

\bibitem{chawla}
N.~V. Chawla, K.~W. Bowyer, L.~O. Hall, and W.~P. Kegelmeyer.
\newblock {SMOTE}: Synthetic minority over-sampling technique.
\newblock {\em Journal of Artificial Intelligence Research}, 16:321--357, jun 2002.

\bibitem{iric}
Bing Zhu, Zihan Gao, Junkai Zhao, and Seppe~K.L.M. {vanden Broucke}.
\newblock Iric: An r library for binary imbalanced classification.
\newblock {\em SoftwareX}, 10:100341, 2019.

\bibitem{wilson}
Dennis~L. Wilson.
\newblock Asymptotic properties of nearest neighbor rules using edited data.
\newblock {\em IEEE Transactions on Systems, Man, and Cybernetics}, SMC-2(3):408--421, 1972.

\bibitem{kaur}
Prabhjot Kaur and Anjana Gosain.
\newblock Empirical assessment of ensemble based approaches to classify imbalanced data in binary classification.
\newblock {\em International Journal of Advanced Computer Science and Applications}, 2019.

\bibitem{rusboost}
Chris Seiffert, Taghi~M. Khoshgoftaar, Jason Van~Hulse, and Amri Napolitano.
\newblock Rusboost: Improving classification performance when training data is skewed.
\newblock pages 1--4, 2008.

\bibitem{ee}
Xu-Ying Liu, Jianxin Wu, and Zhi-Hua Zhou.
\newblock Exploratory undersampling for class-imbalance learning.
\newblock {\em IEEE Transactions on Systems, Man, and Cybernetics, Part B (Cybernetics)}, 39(2):539--550, 2009.

\bibitem{caret}
Max Kuhn.
\newblock Building predictive models in r using the caret package.
\newblock {\em Journal of Statistical Software}, 28(5):1--26, 2008.

\bibitem{ebmc}
Hsiang Hao and {Chen}.
\newblock ebmc: Ensemble-based methods for class imbalance problem.
\newblock 2022.
\newblock R package version 1.0.1.

\bibitem{platt}
Bj{\"o}rn B{\"o}ken.
\newblock On the appropriateness of platt scaling in classifier calibration.
\newblock {\em Information Systems}, 95:101641, 2021.

\bibitem{r}
{R Core Team}.
\newblock {\em R: A Language and Environment for Statistical Computing}.
\newblock R Foundation for Statistical Computing, Vienna, Austria, 2021.

\bibitem{gg}
Hadley Wickham.
\newblock {\em ggplot2: Elegant Graphics for Data Analysis}.
\newblock Springer-Verlag New York, 2016.

\bibitem{epi}
Ewout~W Steyerberg, Andrew~J Vickers, Nancy~R Cook, Thomas Gerds, Mithat Gonen, Nancy Obuchowski, Michael~J Pencina, and Michael~W Kattan.
\newblock Assessing the performance of prediction models: a framework for traditional and novel measures.
\newblock {\em Epidemiology (Cambridge, Mass.)}, 21(1):128--138, 01 2010.

\bibitem{pROC}
Xavier Robin, Natacha Turck, Alexandre Hainard, Natalia Tiberti, Frédérique Lisacek, Jean-Charles Sanchez, and Markus Müller.
\newblock proc: an open-source package for r and s+ to analyze and compare roc curves.
\newblock {\em BMC Bioinformatics}, 12:77, 2011.

\bibitem{1-mimic_iii}
Alistair E.~W. Johnson, Tom~J. Pollard, Lu~Shen, Li-Wei~H. Lehman, Mengling Feng, Mohammad Ghassemi, Benjamin Moody, Peter Szolovits, Leo~Anthony Celi, and Roger~G. Mark.
\newblock {MIMIC}-{III}, a freely accessible critical care database.
\newblock {\em Scientific Data}, 3:160035, May 2016.

\bibitem{2-mimic_iii}
Alistair Johnson, Tom Pollard, and Roger Mark.
\newblock {MIMIC}-{III} clinical database (version 1.4).
\newblock {\em PhysioNet}, 2016.

\bibitem{probst}
Philipp Probst, Marvin~N. Wright, and Anne-Laure Boulesteix.
\newblock Hyperparameters and tuning strategies for random forest.
\newblock {\em {WIREs} Data Mining and Knowledge Discovery}, 9(3), jan 2019.

\bibitem{rose}
Nicola Lunardon, Giovanna Menardi, and Nicola Torelli.
\newblock {ROSE}: a {P}ackage for {B}inary {I}mbalanced {L}earning.
\newblock {\em {R} Journal}, 6(1):82--92, 2014.

\end{thebibliography}

\end{document}